\documentclass[10pt,journal]{IEEEtran}
\usepackage{amsmath}
\usepackage{amssymb}
\usepackage{xcolor}

\usepackage{makecell, multirow}
\usepackage{subcaption, hyperref}
\usepackage{booktabs}
\usepackage{gensymb}

\def\hatx{\mathbf{\hat{x}}}
\def\hatc{\mathbf{\hat{c}}}
\def\x {\mathbf{x}}
\def\c {\mathbf{c}}
\def\y {\mathbf{y}}

\ifCLASSOPTIONcompsoc
  \usepackage[nocompress]{cite}
\else
  \usepackage{cite}
\fi

\ifCLASSINFOpdf
  \usepackage[pdftex]{graphicx}
  \graphicspath{{../pdf/}{../jpeg/}}
   \DeclareGraphicsExtensions{.pdf,.jpeg,.png}
\else
\fi

\usepackage{array}

\usepackage{url}

\begin{document}
%
\title{Variational Deep Image Restoration}
%
%
%
%

\author{Jae Woong Soh,~\IEEEmembership{Student Member,~IEEE,}
        and Nam Ik Cho,~\IEEEmembership{Senior Member,~IEEE}
               
\IEEEcompsocitemizethanks{\IEEEcompsocthanksitem J. W. Soh and N. I. Cho are with the Department
of Electrical and Computer Engineering, Seoul National University, INMC, Seoul,
Korea.
Corresponding: nicho@snu.ac.kr.
This work was supported in part by the National Research Foundation of Korea(NRF) grant funded by the Korea government(MSIT) (2021R1A2C2007220), and in part bynSamsung Electronics Co., Ltd.
}
\thanks{Manuscript received April 19, 2005; revised August 26, 2015.}
}

%
%

\markboth{Journal of \LaTeX\ Class Files,~Vol.~14, No.~8, August~2015}%
{Shell \MakeLowercase{\textit{et al.}}: Bare Demo of IEEEtran.cls for Computer Society Journals}
%



\IEEEtitleabstractindextext{%
\begin{abstract}
This paper presents a new variational inference framework for image restoration and a convolutional neural network (CNN) structure that can solve the restoration problems described by the proposed framework. Earlier CNN-based image restoration methods primarily focused on network architecture design or training strategy with non-blind scenarios where the degradation models are known or assumed. For a step closer to real-world applications, CNNs are also blindly trained with the whole dataset, including diverse degradations. However, the conditional distribution of a high-quality image given a diversely degraded one is too complicated to be learned by a single CNN. Therefore, there have also been some methods that provide additional prior information to train a CNN. Unlike previous approaches, we focus more on the objective of restoration based on the Bayesian perspective and how to reformulate the objective. Specifically, our method relaxes the original posterior inference problem to better manageable sub-problems and thus behaves like a divide-and-conquer scheme. As a result, the proposed framework boosts the performance of several restoration problems compared to the previous ones. Specifically, our method delivers state-of-the-art performance on Gaussian denoising, real-world noise reduction, blind image super-resolution, and JPEG compression artifacts reduction.
\end{abstract}

}

\maketitle

\IEEEdisplaynontitleabstractindextext

%
\IEEEpeerreviewmaketitle

\section{Introduction}\label{sec:introduction}


%
%
%
%
\IEEEPARstart{I}{mage} restoration is an important low-level vision problem because the images are usually corrupted by many kinds of degradation during the imaging and transmission processes. It is typically an ill-posed inverse problem because the image degradation procedures are irreversible. Some examples of degradations are noise corruption, blurring, spatial subsampling, compression artifacts, etc., and there have been many approaches to alleviating or restoring from degradations. Image restoration aims to enhance the visual quality of images for human observability and play the role of preprocessing to boost other high-level vision tasks' performances.

The image degradation is generally modeled as 
\begin{equation}
\y=T(\x)+\mathbf{n},
\end{equation}
where $\y$, $\x$, and $\mathbf{n}$ denote an observed low-quality image, its corresponding high-quality image, and additive noise, respectively. $T(\cdot)$ represents the degradation function, which is usually non-invertible. The overall goal of the image restoration problem can be regarded as to learn the underlying posterior distribution $p(\x|\y)$, and many classical methods estimate the clean image $\x$ based on the maximum-a-posteriori (MAP) inference, with the appropriately designed priors. Traditionally, handcrafted priors have played an important role in this approach, where appropriate domain-relevant and task-relevant priors were designed for the given kind of restoration problem. In the Bayesian perspective, the MAP inference can be divided into likelihood term and prior term as
\begin{align}
\hatx &= \arg \max_\x \log p(\x|\y),\label{eq:MAP}\\
&= \arg \max_\x \log p(\y|\x) + \log p(\x).
\end{align}
The likelihood term is usually modeled as L2-norm based on the Gaussian noise assumption (task-relevant prior), and the prior term is designed based on experience and knowledge about natural images (domain-relevant prior). Some of widely used priors are total variation, sparsity, low-rankness, and non-local self-similarity \cite{tv1, K-SVD, non-local1, non-local2, non-local3, A+, WNNM, BM3D, TWSC, milanfar2012tour}. These strong priors often work well, but explicit constraints on the design limit the performance compared to current data-driven methods.

Recently, CNNs have shown breakthrough performances in overall image restoration problems \cite{DnCNN, FFDNet, NLRN, RNAN, SRCNN, EDSR, IKC, CSNLN, ARCNN, AGARNet} by learning the mapping between $\y$\ and $\x$, {\em i.e.}, by implicitly learning the priors. Most of the early works aimed at non-blind image restoration, or in other words, aimed at specific cases where the degradation information such as noise level for denoising or blur kernel for super-resolution is known. Given the specific degradation models, they generally focused on network architecture design, training datasets, and learning strategies. Afterward, some works focused on more practical solutions such that a single network can cope with diverse degradations \cite{FFDNet, SRMD, ZSSR, MZSR, USRNet, UDVD}, or blind scenarios \cite{ATDNet, IKC, kernelGAN, VDN, DUBD}. Meanwhile, some researchers addressed the problem of the blurry results due to the use of pixel-wise mean squared error (MSE) loss and thus investigated new target objectives based on the generative adversarial net (GAN) \cite{GAN} \cite{SRGAN, SFT-GAN, NatSR, RankSRGAN}.

The above-stated non-blind restoration methods, which we will refer to as ``specific models,'' are not practical because we need to prepare many separate networks to cope with diverse degradation models. When an input image bears a degradation different from the trained one, a severe performance drop occurs due to the domain discrepancy between training and test image distributions. To alleviate this problem, na\"{i}ve blind models are proposed, in which a single network is trained with a training dataset consisted of images having various degradations with broad parameter ranges \cite{DnCNN, MemNet, VDSR, UNLNet}. However, the na\"{i}ve blind model generally performs worse than the non-blind one for a given specific degradation because the conditional distribution $p(\x|\y)$ of the blind training dataset is more diverse and difficult to learn than that of the non-blind one.

Hence, instead of the na\"{i}ve blind training, some methods adopted the two-stage framework for blind image restoration, which is a combination of estimator and restorer. The estimator extracts features related to task-specific prior information, which is fed to the restorer along with the input image \cite{FFDNet, DUBD, AGARNet, CBDNet, SRMD, UDVD}. The restorer is trained to adjust its behavior according to the input and the prior information from the estimator. The prior information in these methods is generally defined based on human knowledge, where the parameters of the degradation are learned by supervised training. For example, they defined the prior knowledge as the spatially varying noise variance map in the case of image denoising.
Based on previous approaches and their results, we hypothesize that providing additional information can be interpreted as dividing a complex distribution into simpler sub-distributions. Thus, it will eventually make the network easier to learn the overall task, further boosting the performance.

Meanwhile, earlier methods trained the networks with synthetic datasets that deviate from the real-world situation. For example, the additive white Gaussian noise (AWGN) model is assumed when constructing datasets for denoising, and the synthetic decimation model with the bicubic blur kernel is used for the single image super-resolution (SISR). Since real-world degradations are different from these assumptions, many researchers also developed CNNs to restore the real-world degraded images \cite{Nam, SIDD, DND, CBDNet, RIDNet, CycleISP, CameraSR, Zoom, RealSR, DRealSR}. The restoration of real-world images has been a challenging task, mainly in two aspects. First, it is hard to build a paired dataset $\{\y_i,\x_i\}_{i=1}^{N}$ for the real-world scenario. Some works assumed degradations with more complex parametric models such as heteroscedastic or Poisson mixture noise \cite{Noise1, Noise2, Noise3, Noise4}. Some others imitated complex degradation based on camera pipeline \cite{CBDNet, UnProcessing, Physics, RawSR}. Instead of model-based methods, several methods learned the degradation based on generative models such as GAN \cite{GCBD, KMSR} or normalizing flow \cite{NoiseFlow}. Some researchers have taken many pairs of real-world images with careful considerations (misalignment, brightness, clipping, etc.) and precise image acquisition settings \cite{SIDD, DND, CameraSR, Zoom, RealSR, DRealSR}. These methods can alleviate the data scarcity problem, but such acquisition processes are costly and labor-intensive.
Second, learning such a complex real-world distribution may be a burdensome task to a single CNN. The distribution of images in real-world situations is very diverse, depending on some criteria, such as exposure, camera manufacturers, capturing environments, etc. Also, the additional prior information such as noise level is unavailable, limiting the training scheme within the na\"{i}ve blind model.

To address the above issues, we propose a new method that can handle blind scenarios, namely Variational Deep Image Restoration (VDIR). Our approach is universal, as it can be adopted for generic image restoration problems. We split the objective of a given image restoration task into simpler sub-problems based on variational approximation, which eventually eases the overall task. Specifically, we provide efficient prior information to the restorer to accurately infer the posterior distribution. For this, we first seek efficient prior information based on the Bayesian perspective and try to approximate posteriors based on our approximated variational distribution.

In summary, our contributions are as follows.
\begin{itemize}
\item We propose an end-to-end trainable CNN, which can be universally applied to blind and real-world image restoration tasks. 
\item Based on variational approximation, we tackle the image restoration problem by relaxing the original problem into easier sub-problems.
\item We seek to find efficient prior information that incorporates both task- and domain-relevant information.
\item We have shown that the proposed method achieves state-of-the-art performances in several image restoration tasks, such as AWGN image denoising, real-noise denoising, blind super-resolution, and JPEG compression artifacts reduction.
\end{itemize}

\section{Related Work}
In this section, we categorize previous CNN-based image restoration methods according to the target objectives. Then, we further discuss prior information that is exploited in previous methods. For the entire paper, $p_{data}(\cdot)$ denotes the underlying data distribution.

\begin{figure}	
	\begin{center}				
		\captionsetup{justification=centering}
		\begin{subfigure}[t]{0.26\linewidth}
			\centering
			\includegraphics[width=1\columnwidth]{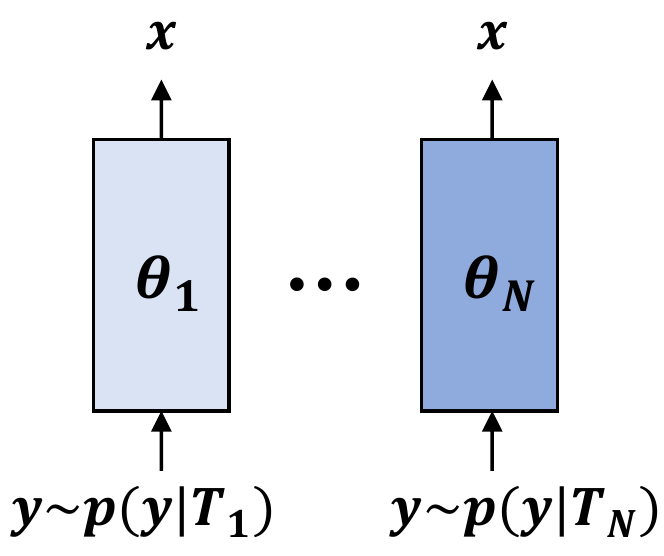}
			\caption{Specific Model}
			\label{fig:specific}
		\end{subfigure}
		\begin{subfigure}[t]{0.18\linewidth}
			\centering
			\includegraphics[width=1\columnwidth]{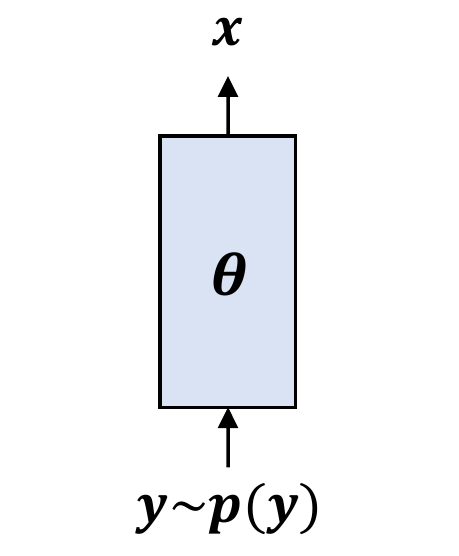}
			\caption{Na\"{i}ve Model}
			\label{fig:naive}
		\end{subfigure}
		\begin{subfigure}[t]{0.25\linewidth}
			\centering
			\includegraphics[width=1\columnwidth]{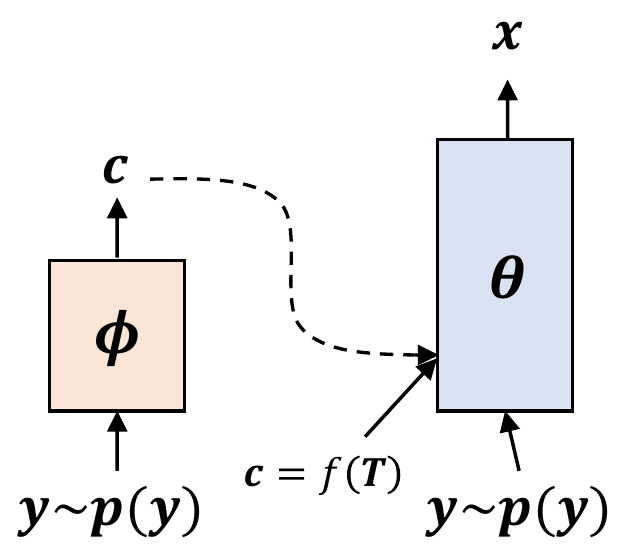}
			\caption{Two-stage Model}
			\label{fig:two-stage}
		\end{subfigure}
		\begin{subfigure}[t]{0.24\linewidth}
			\centering
			\includegraphics[width=1\columnwidth]{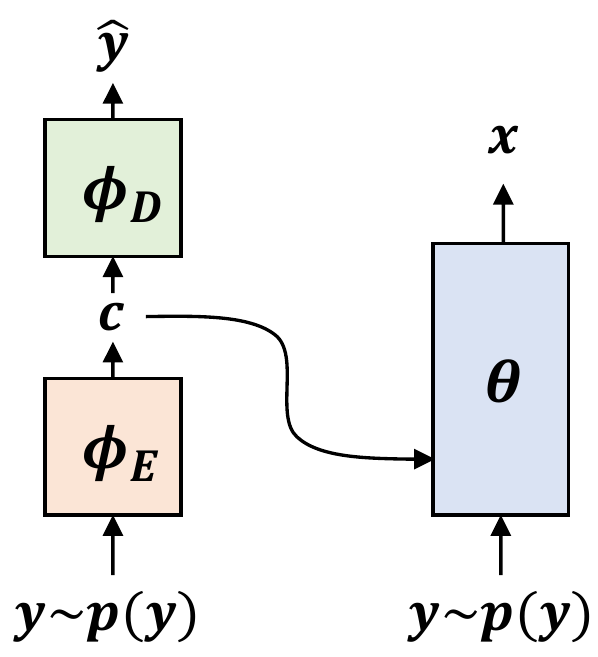}
			\caption{Ours}
			\label{fig:ours}
	\end{subfigure}
	\end{center}
	\caption{Categories of image restoration frameworks.}
	\label{fig:related}
\end{figure}

\subsection{Specific Non-blind Model}
The early CNN-based image restoration methods \cite{DnCNN, SRCNN, NLRN, EDSR, RCAN, SRFBN, ARCNN} adopted ``specific non-blind model,'' which is a separate network parameterized by $\theta_i$ and trained for a specific degradation $T_i$, as shown in \figurename{~\ref{fig:specific}}. Its target objective can be expressed as
\begin{equation}
\hat{\theta_i}=\arg \max_{\theta_i} \mathbb{E}_{p_{data}(\x, \y, \c)}[\log p_{\theta_i}(\x|\y,\c_i)]
\end{equation}
where $\c_i$ is a function of $T_i$, \emph{e.g.}, the noise level for denoising, blur kernels or scaling factors for super-resolution, the quality factor for compression artifacts reduction, etc. The non-blind approaches have limitations in practical use because we need to know or accurately estimate the degradation information, which is generally not an easy problem. Also, we need to prepare many networks in the bag to cope with large variations of degradation in the real world, requiring too large memories.

\subsection{Na\"{i}ve Blind Model}
Most of the blind methods are in the category of na\"{i}ve blind model, which rely only on the representation power of CNN, as described in \figurename{~\ref{fig:naive}}. For example with denoising, since there is no prior information such as noise level in real-world noisy images, it would be an appropriate approach to adopt the na\"{i}ve blind model. The target objective is expressed as
\begin{equation}
\hat{\theta}=\arg \max_{\theta} \mathbb{E}_{p_{data}(\x, \y)}[\log p_{\theta}(\x|\y)],
\end{equation}
for training a single network parameterized by $\theta$ to capture the conditional distribution. In general, na\"{i}ve blind models show worse performance than the specific ones \cite{DnCNN, VDSR, OneSize}, since the marginal posterior $p(\x|\y)$ is more complicated than the posterior given the degradation, $p(\x|\y, T)$. Specifically, several methods such as \cite{DnCNN, UNLNet, GCBD, RIDNet, KMSR, Meta-SR} come under this category.

\subsection{Two-Stage Blind Model}
Some recent methods can be categorized as a two-stage blind model, where the degradation parameters $\c$, which is a function of $T$, are first estimated and then fed to the restoration network along with the input image. Its target objective is
\begin{align}
\hat{\phi} =& \arg \max_{\phi} \mathbb{E}_{p_{data}(\y, \c)}[ \log p_{\phi}(\c|\y)],\\
\hat{\theta} =&\arg \max_{\theta} \mathbb{E}_{p_{data}(\x, \y, \c)} [\log p_{\theta}(\x|\y,\c)]],
\end{align}
where $\phi$ and $\theta$ denote the parameters of the estimator and the restorer, respectively. For the Gaussian denoising, $\c$ is selected as the standard deviation of Gaussian distribution \cite{FFDNet, ATDNet, DUBD}, and for the real-noise, more complicated parameters are selected \cite{CBDNet, VDN, AINDNet}. For super-resolution, blur kernel is selected \cite{SRMD, ZSSR, MZSR, IKC}, and quality factor is selected for JPEG compression artifacts reduction \cite{AGARNet}. Note that prior information on the degradation model is additionally required for this setting.

\subsection{Additional Prior Information}
For developing deep learning-based image/video restoration algorithms, many kinds of prior information have been considered to better help the network learn such complicated problems. Mainly, the task-relevant priors and the domain-relevant priors are considered.

Typically, the task-relevant priors such as the degradation model with its parameters are widely used \cite{FFDNet, DUBD, IKC, AGARNet}.
On the other hand, the domain-relevant priors, which have long been considered from traditional image restoration, have also been considered. For example, natural image priors such as non-local self-similarity \cite{BMCNN, NLRN, CSNLN} or sparsity by wavelet transforms \cite{DWSR, MWCNN, WaveletSRNet} are additionally provided to the network or incorporated in a somewhat different way by conditioning the network architectures. Semantic information is also utilized to guide the rebuilding of high-frequency details in super-resolution tasks \cite{SFT-GAN}. Even more, the facial domain-specific priors are used for the face-specified super-resolution. For example, facial landmarks or heatmaps are exploited for face super-resolution in \cite{SuperFAN, FSRNet}.
For video restoration problems, the main domain-relevant prior may be temporal coherency, and some of the methods use optical-flow as the domain-relevant prior \cite{ARTN, VSRNet, FRVSR, ToFlow, TDAN}. Note that these approaches aim at specific tasks which cannot be applied to generic image restoration problems, and the pre-designed priors are often sub-optimal \cite{ToFlow}.

Besides, there have been other priors exploited for image restoration tasks. It has been noticed that residual networks, which bypass diverse levels of information within the network, can be considered a way of providing additional priors to each inner block of the network \cite{DuRN}. However, it is mainly related to architectural design, and any constraints on the residual information are imposed while training. Very recently, deep generative priors based on pre-trained GAN models \cite{StyleGAN, StyleGAN2} have been proposed in \cite{mGANprior, GLEAN, DGP}. The prior information encapsulated in a pre-trained GAN shows dramatic improvements in image restoration tasks, but their performance gains mostly rely on the pre-trained networks and external dataset. Moreover, most pre-trained GANs (StyleGAN \cite{StyleGAN}) are trained with datasets in a specific category such as face or cat. Therefore, available image categories are limited. Also, auxiliary supervision on residual image works as a prior for dynamic scene deblurring \cite{Meta-Auxiliary}. However, the prior is a reconstruction of the residual, and the essential part of the above-stated work is a meta-learning step that requires gradient descent updates for inference.
On the other hand, our VDIR does not require any supervision on the external dataset or the pre-trained network, and the prior is jointly learned while training. Also, only a feed-forward pass is required to tune the behavior of the restoration network.

\section{Variational Deep Image Restoration}
This section presents our target objective regarding the inference of the high-quality image $\x$ and the latent prior information $\c$, then reformulates the problem to tractable sub-problems.

\subsection{Problem Statement}
For a given low-quality image $\mathbf{y}$, the objective is to find a latent high-quality image $\mathbf{x}$, which is described as a MAP inference problem in eq.~(\ref{eq:MAP}). To solve this problem, we start from a different perspective compared to former MAP frameworks.

Let us first consider the joint distribution $p(\x,\y)$ and describe a framework that generates the low and high-quality images.
Our goal is to learn underlying joint distribution and further infer the posterior distribution $p(\x|\y)$ from the joint distribution.
For this, we introduce a new latent random variable $\c$ in this framework, which implies both domain- and task-relevant properties.
In our framework, we assume that the low-quality images are generated based on a procedure involving $\c$, which can be split into three processes:
(1) given a domain prior $\kappa$ and a degradation $T$, the latent variable $\c$ is generated from some conditional distribution $p(\c;\kappa, T)$,
(2) from $\c$ and a natural image prior $\xi$, a high-quality natural image is generated from $p(\x|\c;\xi)$, and
(3) from $\x$ and $\c$, a low-quality degraded image is generated from $p(\y|\x,\c)$.

In this framework, we bring a new inference problem of the posterior $p(\c|\x, \y)$, where the latent $\c$ includes both domain- and task-relevant information through the given information of high- and low-quality images. However, this inference problem is intractable. Also, our other objective is to infer $\x$, which cannot be observed during the inference.

In summary, our inference problems of interest are 
\begin{itemize}
	\item the inference of $\log p(\c|\x,\y)$, and
	\item the inference of $\log p(\x|\y)$,
\end{itemize}
which are approximated and solved by training CNNs with appropriate datasets for the given inference problems.

\subsection{Proposed Variational Lower Bound}
To approximate the posterior $p(\c|\x, \y)$, we introduce a tractable probability distribution $q(\c|\y)$.
Then, the joint probability distribution $\log p(\x,\y)$ can be reformulated as
\begin{align}\nonumber
\log p(\x, \y) =& \mathbb{E}_{\c \sim q(\c|\y)} [\log p(\x|\y,\c)]\\\nonumber
& + D_{KL}(q(\c|\y)||p(\c|\x,\y))\\\nonumber
& - D_{KL}(q(\c|\y)||p(\c))\\
& + \mathbb{E}_{\c \sim q(\c|\y)} [\log p(\y|\c)], \label{eq:var0}
\end{align}
with some prior distribution $p(\c)$. The details of the derivation can be found in the {\em supplemental material}.
To approximate the intractable KL divergence term between $q(\c|\y)$ and $p(\c|\x, \y)$, we introduce a \emph{variational lower bound} $\mathcal{L}$.

\vspace{0.2cm}
\noindent\textbf{Definition 1}
\emph{Variational lower bound} $\mathcal{L}$ is defined as
\begin{align}\nonumber
\mathcal{L} =& \mathbb{E}_{\c \sim q(\c|\y)} [\log p(\x|\y,\c)]\\
& - D_{KL}(q(\c|\y)||p(\c)) + \mathbb{E}_{\c \sim q(\c|\y)} [\log p(\y|\c)].
\label{eq:var1}
\end{align}

\noindent\textbf{Theorem 1}
\textit{Given a low-quality image $\y$ and its underlying high-quality image $\x$, the joint log-distribution $\log p(\x,\y)$ can be reformulated including variational lower bound $\mathcal{L}$ as}
\begin{equation}
\log p(\x,\y) = \mathcal{L} + D_{KL}(q(\c|\y)||p(\c|\x,\y)).
\label{eq:var2}
\end{equation}
Then,
\begin{equation}
\log p(\x,\y) \geq \mathcal{L}. \label{eq:lowerbound}
\end{equation}
The proof of eq.~(\ref{eq:var2}) is straightforward from eqs. (\ref{eq:var0}) and (\ref{eq:var1}). Also, eq.~(\ref{eq:lowerbound}) comes from the fact that the KL divergence in eq.~(\ref{eq:var2}) is non-negative.

\vspace{0.2cm}
\noindent\textbf{Definition 2}
We also define a log-posterior $q(\x|\y)$ which approximates the original posterior $p(\x|\y)$ as
\begin{align}\nonumber
q(\x|\y) &= \int_\c q(\x,\c|\y) d\c\\
&= \int_\c p(\x|\y,\c)q(\c|\y) d\c.
\end{align}
Then, the MAP inference given $\y$ can be done by $\arg \max_\x \log q(\x|\y)$.

Through our reformulation, maximizing the likelihood of the joint probability distribution is approximated as to maximize our \emph{variational lower bound}.
Notably, the first term of $\mathcal{L}$ in eq.~(\ref{eq:var1}) is the only term relevant to the relation between $\x$ and $\y$, responsible for image restoration.
The other terms are regularization terms, which impose constraints on the latent variable $\c$. The second term is the KL divergence, which constrains the latent distribution, and the third term is the auto-encoder reconstruction term of the low-quality images.

As neural networks are experts at inference, in an amortized way \cite{amortized, AVI}, we employ three CNNs parameterized by $\theta$, $\phi_E$, and $\phi_D$ for the variational inference. Specifically, our final objective is
\begin{align}
\arg \max_{\theta, \phi_E, \phi_D} \mathbb{E}_{p_{data}(\x, \y)}[\nonumber
&\mathbb{E}_{\c \sim q_{\phi_E}(\c|\y)} [\log p_{\theta}(\x|\y,\c)]\\\nonumber
& - D_{KL}(q_{\phi_E}(\c|\y)||p(\c))\\
& + \mathbb{E}_{\c \sim q_{\phi_E}(\c|\y)} [\log p_{\phi_D}(\y|\c)]],
\end{align}
with underlying empirical data distribution $p_{data}(\x, \y)$.
By introducing such regularizations, our restorer can approximately solve a MAP problem according to the latent $\c$, where $\c$ is the variable involved in the low-quality image generation process, which involves both task- and domain-relevant prior information. In other words, our restorer divides the problem according to the latent vector $\c$, where $\c$ should ``imply'' the low-quality image manifold.

\subsection{Corresponding Loss Terms}
\noindent\textbf{First Term}
The first term means to estimate $\x$ conditioned on $\y$ and $\c$. If there is a function (network) that performs this as $\hat{\x}=f(\y,\c)$, then the problem is to train the network to minimize the difference between $\x$ and $\hat{\x}$.
We adopt L1 loss between the ground-truth clean image and the inferred output to minimize the distortion based on the i.i.d. Laplacian prior. We denote the corresponding loss term as $\mathcal{L}_{res}$:
\begin{equation}
\mathcal{L}_{res}= ||\x-\hat{\x}||_1,
\end{equation}
where $\hat{\x}=f(\y,\c)$ as stated above, and $f(\y,\c)$ is our main restoration network that will be explained in Sec. IV.E, with Fig. 3.

\noindent\textbf{Second Term}
The second term measures the fidelity of $\c$ extracted through the encoder $\phi_E$ with the input $\y$, and this is represented as the KL divergence between $q_{\phi_E}(\c|\y)$ and $p(\c)$.
The KL divergence between the prior distribution and the posterior can be calculated analytically.
Since $p(\c)$ is a modeling of the latent distribution, any parametric distribution model can be used \cite{VAE, BetaVAE}, and we use a standard Gaussian for simplicity and convenience.
Specifically, we set the prior $p(\c)$ as Gaussian distribution with zero mean and unit covariance. Thus, the KL divergence term is

\begin{align}\nonumber
D_{KL}(\mathcal{N}(\mu&, \Sigma)||\mathcal{N}(0, I)) \\
= \frac{1}{2}[-\log &\det (\Sigma) - n + tr (\Sigma) + \mu^T \mu],\\\nonumber
D_{KL}(q(\c|\y)||p(\c))&= \frac{1}{2}[ \sum_{i=1}^{n} (-\log \sigma_i^2 -1 + \sigma_i^2 + \mu_i^2)],
\end{align}
where $n$ is the dimension of the random variable.

\noindent\textbf{Third Term}
In the case of the third term, which is the auto-encoder reconstruction term, we may also adopt L1 loss between the noisy image $\y$ and the decoder’s output $\hat{\y}$, like in the first term. However, using only pixel-wise loss strictly assumes $p(\y|\c)$ to be a family of the probability distribution of i.i.d. Laplacian or Gaussian. Hence, we additionally adopt adversarial loss \cite{GAN} to relax and better learn the low-quality image distribution. Specifically, we adopt non-saturating GAN loss \cite{GAN}, corresponding to minimizing Jensen-Shannon divergence between $p_{data}(\y)$ and $p_{\phi_{D}}(\y|\c)$.
It is expressed as
\begin{align}
\mathcal{L}_{D}&= - \log D(\y) - \log(1-D(\hat{\y})),\\
\mathcal{L}_{adv}&= - \log D(\hat{\y}) 
\end{align}
where $\mathcal{L}_{D}$ and $\mathcal{L}_{adv}$ denote the discriminator loss and the adversarial loss, respectively.
For each specific task, we may inject known priors if available. In this case, we add an estimation loss so that the latent space learns the known degradation information.
Eventually, the latent $\c$ works as an additional prior for the restorer, which contains an abstract of the low-quality image distribution.

With the above notions, the loss $\mathcal{L}_{rec}$ for the third term is described as
\begin{equation}
\mathcal{L}_{rec}= ||\y-\hat{\y}||_1 + \lambda_1 \mathcal{L}_{adv} + \lambda_2 ||f(T) - EST(\c)||_1,
\end{equation}
where $EST(\cdot)$ is a simple two-layer CNN (conv-relu-conv).
$f(T)$ is an injected prior knowledge when available. $T$ and $f(T)$ are symbolic expressions, where $T$ stands for a kind of degradation and $f(T)$ is defined differently depending on the restoration tasks. In the following sections, we will show examples of $f(T)$ for Gaussian denoising, super-resolution, and compression artifacts reduction. When we do not have the knowledge, we do not use it by setting $\lambda_2=0$.

\noindent\textbf{Overall Loss}
The overall loss is the sum of three terms where normalized KL divergence is used and multiplied by $\beta$ \cite{BetaVAE}.
Then, the final objective is
\begin{equation}
\arg \min_{\theta, \phi_E, \phi_D}  \mathbb{E}_{p_{data}(\x, \y)}[\mathcal{L}_{res} + \beta  D_{KL} + \mathcal{L}_{rec}].
\end{equation}

\section{Discussions}
\subsection{Difference from VDN}
Both our VDIR and VDN \cite{VDN} adopt variational framework for image denoising or image restoration problem. However, there are mainly three differences. First, VDN models the noise distribution with non-i.i.d Gaussian distribution, but ours does not explicitly model the distribution, instead it is implicitly learned through the network. Second, VDN sets the latent variable as latent clean image $\mathbf{z}$ and approximates the distribution over the noisy observation $\y$: $\log p(\y)$. On the other hand, our method introduces the latent variable $\c$ containing beneficial information and approximates the joint distribution of both $\x$ and $\y$: $\log p(\x, \y)$. Lastly, VDN analytically solves the likelihood function, whereas our method uses a neural network to approximate likelihood.

\subsection{Difference from Plug-and-play based methods}
Diverse plug-and-play based methods also utilize priors for image restoration \cite{ren2019simultaneous, zhang2019deep, liu2020investigating, USRNet, zhang2021plug}. However, there are two main differences compared to ours.
First, the approach and modeling of the target objective function are different. Plug-and-play methods start their formulation based on the MAP framework and explicitly divide the posterior term into the data fidelity term and the prior term. Then, they carefully deal with each term. On the other hand, our method directly learns the posterior based on the variational approximation with implicit prior. Second, the plug-and-play methods are iterative due to their inherent nature of optimization. Although the priors are learned from the dataset, several iterations are required to obtain desirable results. On the contrary, our method requires one feed-forward pass to obtain restored results, which is efficient regarding the computational complexity.

\subsection{Probabilistic View}
In this subsection, we discuss our method with other frameworks with the probabilistic perspective.
Let us assume the restoration loss term as L2-norm for simplicity, which assumes i.i.d. Gaussian distribution as target probabilistic family (L1-norm is associated with Laplacian distribution). We refer a neural network parameterized by $\theta$ as $f_{\theta}(\cdot)$.

The na\"{i}ve blind model which models $p_{\theta}(\x|\y) = \mathcal{N}(f_{\theta}(\y),I/2)$, and learning the data distribution under this family hinders expressiveness. Since the posterior distribution $p(\x|\y)$ including diverse degradation is too complex to be captured by a single Gaussian, its performance cannot be expected as much as the specific model with simpler $p(\x|\y)$ with a single degradation.

On the other hand, our framework models $p(\x|\y, \c) = \mathcal{N}(f_{\theta}(\y, \c), I/2)$. It is still Gaussian but learns different mean values with respect to $\c$, which grants more representation power for learning multiple Gaussians in accordance with $\c$. (Analogous to the relation between Gaussian and Gaussian mixture model.)
Then, the marginal posterior has more representative powers $p(\x|\y) = \int_{\c} p(\x|\y,\c)p(\c|\y) d\c$.
Though, our inference approximates the marginal posterior through Monte-Carlo using only one sample of $\c$.

The two-stage model can be considered a special case of our method where the $q(\c|\y)$ is chosen as deterministic.
In this case, the $c$ is determined based on the prior knowledge and carefully modeled by ``understanding the data.''
Then, the point estimate of $\hatc = \arg \max_\c q(\c|\y)$ is used for the second step inference \cite{DUBD}.
Note that this bi-level optimization scheme would be sub-optimal to the task objective, compared to the joint optimization.
Unlike the two-stage model, our method is more Bayesian and implicitly learns $\c$, which is enforced to contain the degradation information along with the original image content information. In other words, our method conducts Bayesian inference, whereas the two-stage model conducts deterministic estimates. Also, the additional information $\c$ is learned by enforcing the network to ``understand the data.''

\begin{figure}	
	\begin{center}				
		\centering
		\includegraphics[width=0.6\columnwidth]{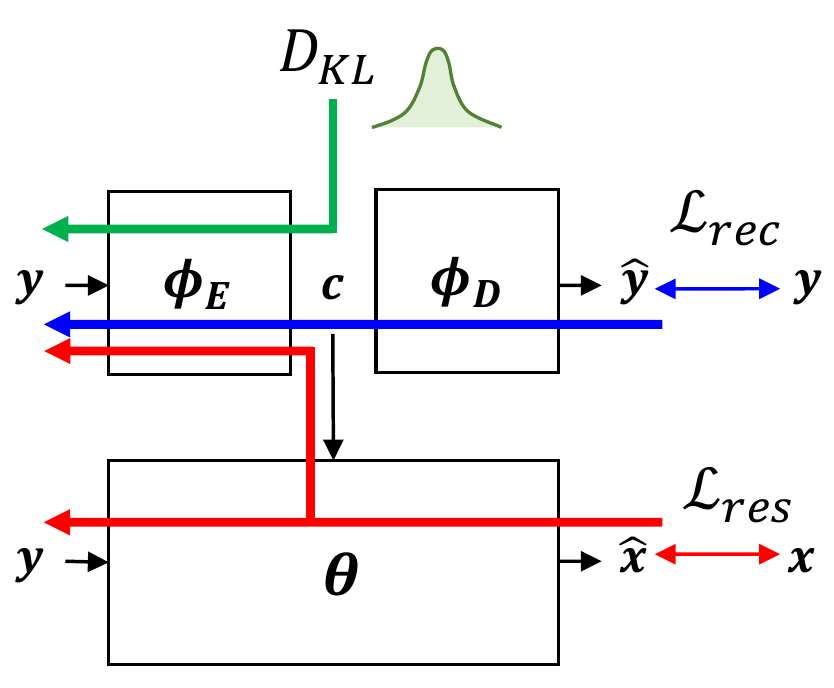}
	\end{center}
	\caption{Overall gradient flows of the loss terms.}
	\label{fig:loss}
\end{figure}

\subsection{Discussions on Loss Terms}
When we use only the first term, it is just a na\"{i}ve approach to solve a blind restoration problem, totally relying on the discriminative power of CNN. In this case, the objective is
\begin{equation}
\arg \max_{\theta, \phi_E} \mathbb{E}_{p_{data}(\x, \y)}[\mathbb{E}_{\c \sim q_{\phi_E}(\c|\y)} [\log p_{\theta}(\x|\y,\c)]].
\end{equation}
Note that no matter what latent distribution $q_{\phi_E}(\c|\y)$ we choose, this criteria is maximized if for each $\c$, $\mathbb{E}_{p_{data}(\x,\y)}[\log p_\theta(\x|\y,\c)]$ is maximized. In other words, there is a trivial solution ``independent'' to $\c$, if our model has the optimal parameter satisfying $\theta^*=\arg \max_\theta \mathbb{E}_{p_{data}(\x, \y)}[\log p_\theta(\x|\y)]$. In this case, $\c$ collapses and the original problem cannot be divided into sub-problems.
The second term gives regularization constraints. The KL divergence term forces the disentanglement of $\c$, giving {\em discriminative power} over observed images $\y$.
The third term further gives constraints on the latent variable $\c$. As the auto-encoder reconstruction term forces the reconstruction from $\c$ to $\y$, this term forces $\c$ to include the information of the degraded image.
Importantly, as in \figurename{~\ref{fig:loss}} that shows the gradient flow of each loss term, the encoder learns not only the reconstruction of the low-quality image (\textcolor{blue}{blue flow}) but also the information suitable for image restoration (\textcolor{red}{red flow}) regularized to a specific probabilistic model through KL divergence (\textcolor{green}{green flow}).
Also, it is trained in an end-to-end manner to let all the information be well reflected to the latent variable.

\begin{table*}[!t]
	\caption{The average PSNR on AWGN denoising. The best results are highlighted in \textcolor{red}{red} and the second best in \textcolor{blue}{blue}.}
	\begin{center}
		\resizebox{0.98\linewidth}{!}{
			\begin{tabular}{c|c|c|c|c|c|c|c}
				\Xhline{4\arrayrulewidth}
				\rule[-1ex]{0pt}{3.5ex}
				Noise level & Dataset & CBM3D \cite{BM3D} & CDnCNN \cite{DnCNN} & FFDNet \cite{FFDNet} & UNLNet \cite{UNLNet} & VDN \cite{VDN} & VDIR (Ours)\\
				\hline\hline
				
				\rule[-1ex]{0pt}{3.5ex}
				\multirow{3}{*}{$\sigma = 10$}&CBSD68&35.91&36.13&36.14& 36.20 & \textcolor{blue}{36.29} & \textcolor{red}{36.34}\\
				\rule[-1ex]{0pt}{3.5ex}
				&Kodak24  & 36.43 & 36.46 & 36.69 & - & \textcolor{blue}{36.85} & \textcolor{red}{37.02}\\
				\rule[-1ex]{0pt}{3.5ex}
				&Urban100& 36.00 &  34.61 & 35.78 & - & \textcolor{blue}{35.97} & \textcolor{red}{36.30} \\
				\hline\hline
				
				\rule[-1ex]{0pt}{3.5ex}
				\multirow{3}{*}{$\sigma = 30$}& CBSD68 & 29.73 & 30.34 & 30.32 & 30.21 & \textcolor{red}{30.64} & \textcolor{red}{30.64} \\
				\rule[-1ex]{0pt}{3.5ex}
				&Kodak24  & 30.75 & 31.17 & 31.27 & 31.18 & \textcolor{blue}{31.67} & \textcolor{red}{31.74} \\
				\rule[-1ex]{0pt}{3.5ex}
				&Urban100& 30.36& 30.00 & 30.53 & 30.41 & \textcolor{blue}{31.14} & \textcolor{red}{31.41}\\
				\hline\hline
				
				\rule[-1ex]{0pt}{3.5ex}
				\multirow{3}{*}{$\sigma = 50$}&CBSD68& 27.38& 27.95 & 27.97 & 27.85 & \textcolor{red}{28.33} & \textcolor{red}{28.33} \\
				\rule[-1ex]{0pt}{3.5ex}
				&Kodak24  & 28.46  & 28.83 & 28.98 & 28.86 & \textcolor{blue}{29.44} & \textcolor{red}{29.49} \\
				\rule[-1ex]{0pt}{3.5ex}
				&Urban100& 27.94 & 27.59 & 28.05 & 27.95 & \textcolor{blue}{28.86} &  \textcolor{red}{29.10}\\
				\hline\hline
				
				\rule[-1ex]{0pt}{3.5ex}
				\multirow{3}{*}{$\sigma = 70$}&CBSD68& 26.00  & 25.66 & 26.55 & - & \textcolor{blue}{26.93} & \textcolor{red}{26.94}\\
				\rule[-1ex]{0pt}{3.5ex}
				&Kodak24  & 27.09  & 26.36 & 27.56 & - & \textcolor{blue}{28.05} & \textcolor{red}{28.10} \\
				\rule[-1ex]{0pt}{3.5ex}
				&Urban100& 26.31  &  25.24 & 26.40 & - & \textcolor{blue}{27.31} & \textcolor{red}{27.55} \\
				
				\Xhline{4\arrayrulewidth}
			\end{tabular}
		}
	\end{center}	
	\label{tab:AWGN}
\end{table*}

\subsection{Network Architecture}

\begin{figure}	
	\begin{center}				
		\captionsetup{justification=centering}
		\begin{subfigure}[t]{0.9\linewidth}
			\centering
			\includegraphics[width=1\columnwidth]{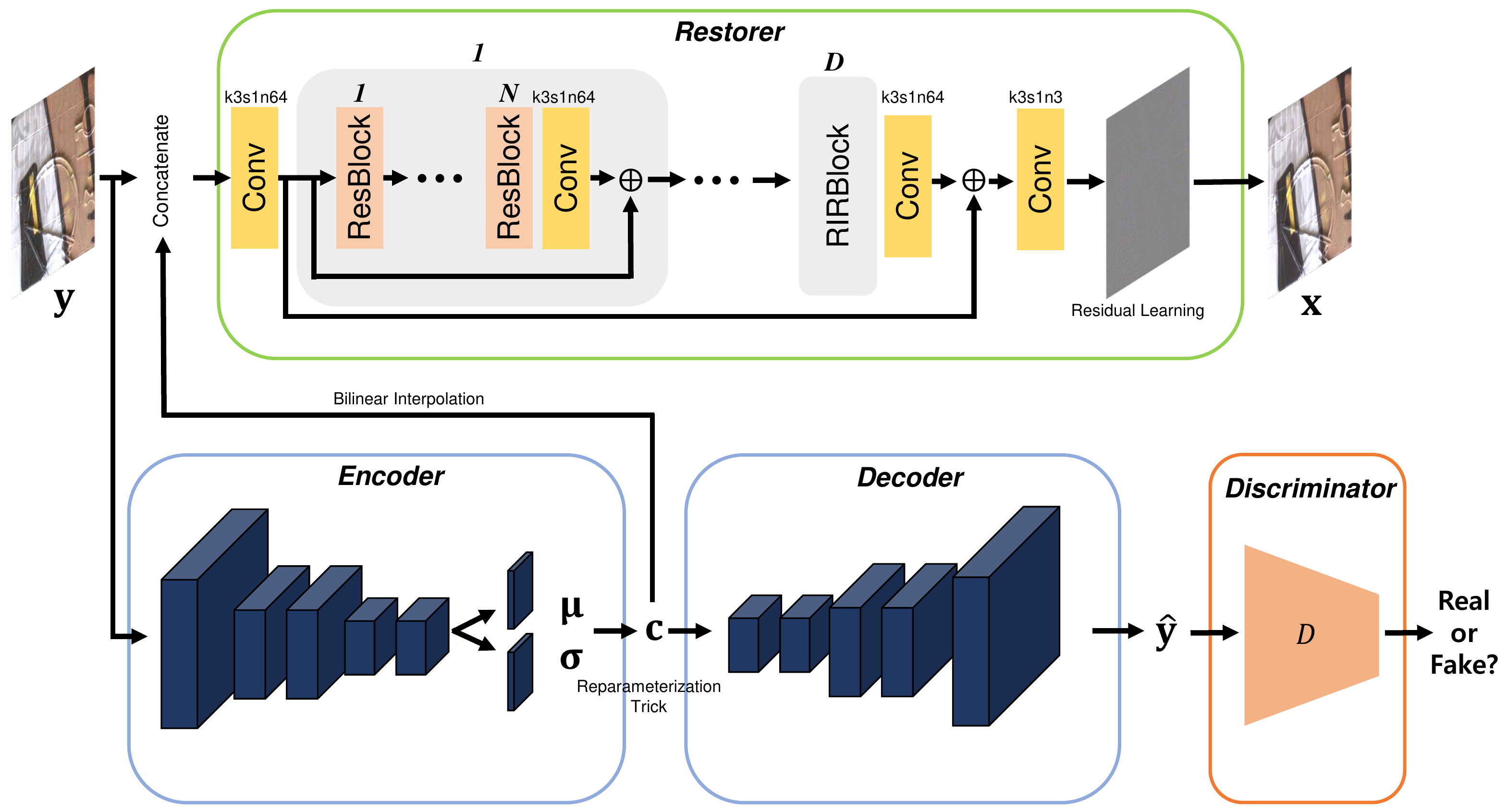}
			\caption{Image Restoration Network}
			\label{fig:restorer}
		\end{subfigure}
		\begin{subfigure}[t]{0.9\linewidth}
			\centering
			\includegraphics[width=1\columnwidth]{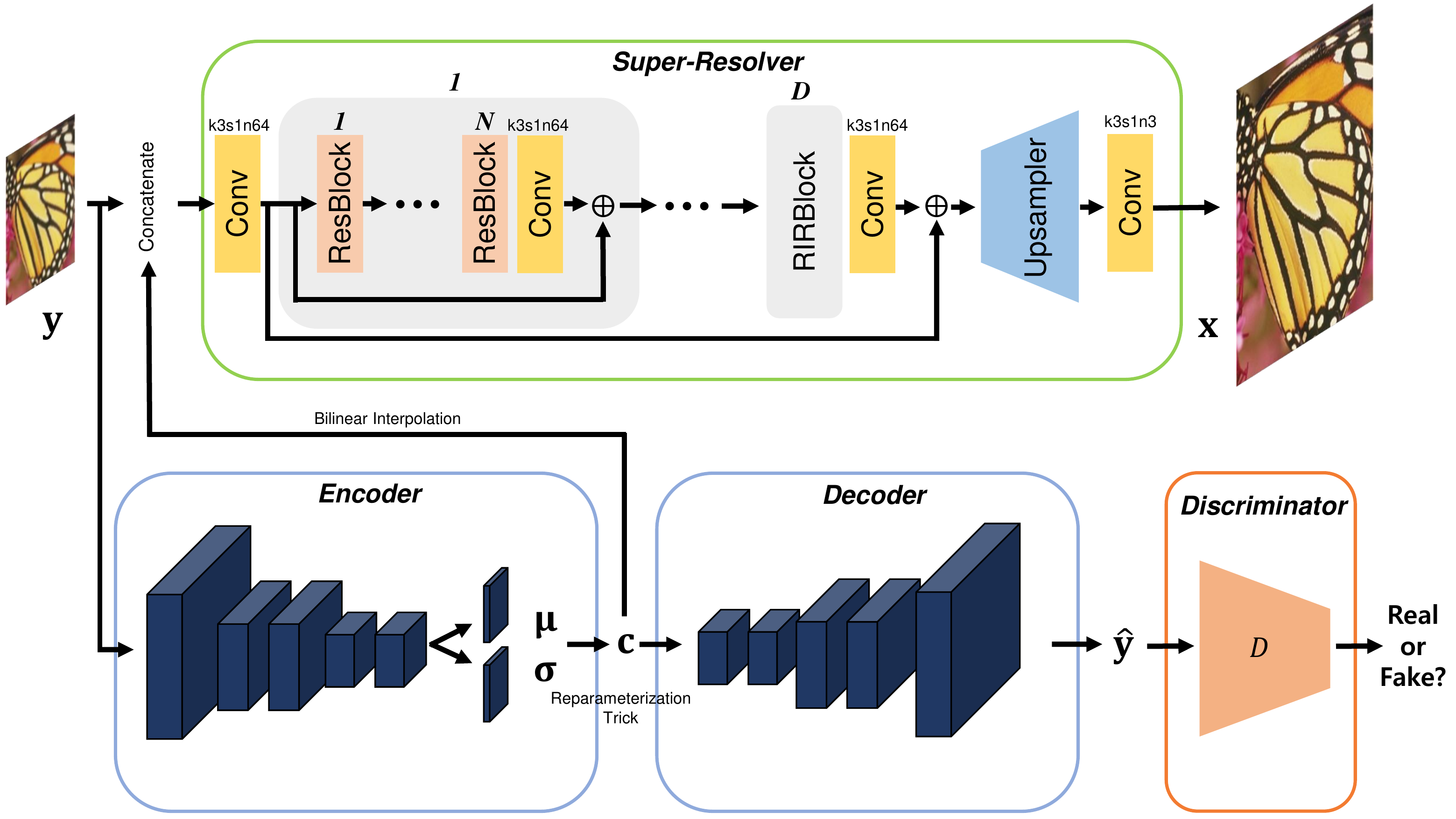}
			\caption{Super-Resolution Network}
			\label{fig:super-resolver}
		\end{subfigure}
	\end{center}
	\caption{The overall architecture of proposed VDIR, where $k$, $s$, and $n$ denote kernel size, stride, and the number of filters, respectively. Image restoration network is adopted for the image denoising and compression artifacts reduction tasks, while the super-resolution network is employed for a super-resolution task. Note that the decoder is not used for the inference. The structure of the discriminator is shown in the supplementary material.}
\label{fig:network}
\end{figure}

The overall network architecture is shown in \figurename{~\ref{fig:network}}. We present two architectures for the generic image restoration and the super-resolution, where the latter is slightly modified from the former. The restorer takes a low-quality image $\y$ with the latent variable $\c$ concatenated along the channel axis to infer the high-quality image $\x$. The restorer is fully convolutional, thus highly scalable. For the restorer, the residual block (ResBlock) is adopted as the basic building block \cite{ResNet, EDSR}. Precisely, the same residual block of \cite{EDSR} is adopted, which consists of a $3\times 3$ convolution layer with $64$ filters followed by the rectified linear unit (ReLU) and another convolution layer (Conv-ReLU-Conv). Then, the input is added to the output of the convolution layer, which forms the skip-connection. The $N$ residual blocks and one convolution layer compose the residual-in-residual block (RIRBlock) \cite{RCAN}. The restorer consists of $D$ RIRBlocks with some convolution layers and a long skip-connection, as shown in \figurename{~\ref{fig:restorer}}. The last convolution layer infers the residual image instead of the high-quality image itself, according to \cite{DnCNN}. The latter part of the restorer is slightly modified for super-resolver. As the spatial feature size is different between $\x$ and $\y$ for the super-resolution, instead of the residual learning, we adopt post-upsampler at the latter part as shown in \figurename{~\ref{fig:super-resolver}}. The upsampler is a sub-pixel convolution layer that consists of a convolution layer followed by a sub-pixel shuffler \cite{ESPCN}.

The encoder and decoder networks are simple feedforward convolutional networks without skip-connection. The encoder decreases the feature map's spatial size twice (one-fourth of its height and width), and the output $\c$ has four channels. For the differentiable Monte Carlo, we adopt the reparameterization trick \cite{VAE, BetaVAE}. The decoder network has symmetrical architecture as the encoder. The details of the encoder, decoder, and discriminator architectures are presented in the {\em supplemental material}.

\begin{table}
	\caption{Results on SIDD \cite{SIDD} benchmark. The best results are highlighted in \textcolor{red}{red} and the second best in \textcolor{blue}{blue}.}
	\label{tab:SIDD}
	\centering
	\resizebox{1.0\linewidth}{!}{
		\begin{tabular}{lcccc}
			\Xhline{4\arrayrulewidth}
			\rule[-1ex]{0pt}{3.5ex}
			Method & Blind/Non-blind & Parameters & PSNR & SSIM\\
			\hline\hline
			\rule[-1ex]{0pt}{3.5ex}
			BM3D \cite{BM3D} & Non-blind & - & 25.65 & 0.685 \\
			\rule[-1ex]{0pt}{3.5ex}
			WNNM \cite{WNNM} & Non-blind & - & 25.78 & 0.809 \\
			\rule[-1ex]{0pt}{3.5ex}
			DnCNN \cite{DnCNN} & Non-blind & 668 K & 23.66 & 0.583 \\
			\rule[-1ex]{0pt}{3.5ex}
			TNRD \cite{TNRD} & Non-blind & 27 K & 24.73 & 0.643 \\
			\rule[-1ex]{0pt}{3.5ex}
			CBDNet \cite{CBDNet} & Blind & 4.4 M & 33.28 & 0.868 \\
			\rule[-1ex]{0pt}{3.5ex}
			RIDNet \cite{RIDNet} & Blind & 1.5 M & 38.71 & 0.914 \\
			\rule[-1ex]{0pt}{3.5ex}
			VDN \cite{VDN} & Blind & 7.8 M & 39.26 & 0.955 \\
			\rule[-1ex]{0pt}{3.5ex}
			AINDNet+TF \cite{AINDNet} & Blind & 13.7 M & 38.95 & 0.952 \\
						\rule[-1ex]{0pt}{3.5ex}
			MIRNet \cite{MIRNet} & Blind & 31.8 M & \textcolor{red}{39.72} & \textcolor{red}{0.959} \\
			\rule[-1ex]{0pt}{3.5ex}
			MPRNet \cite{zamir2021multi} & Blind & 15.7 M & \textcolor{blue}{39.71} & \textcolor{blue}{0.958} \\
			\rule[-1ex]{0pt}{3.5ex}
			InvDN \cite{InvDN} & Blind & 2.6 M & 39.28 & 0.955 \\
			\hline
			\rule[-1ex]{0pt}{3.5ex}
			VDIR (Ours)& Blind & 2.2 M & 39.26 & 0.955 \\
			\rule[-1ex]{0pt}{3.5ex}
			VDIR+ (Ours)& Blind & 2.2 M & 39.33 & 0.956\\
			\Xhline{4\arrayrulewidth}
		\end{tabular}
	}
\end{table}

\begin{table}
	\caption{Results on DND \cite{DND} benchmark. The best results are highlighted in \textbf{bold}.}
	\label{tab:DND}
	\centering
	\resizebox{1.0\linewidth}{!}{
		\begin{tabular}{l|cccc}
			\Xhline{4\arrayrulewidth}
			\rule[-1ex]{0pt}{3.5ex}		
			Method & Blind/Non-blind & Parameters & PSNR & SSIM \\
			\hline\hline
			\rule[-1ex]{0pt}{3.5ex}
			BM3D \cite{BM3D} & Non-blind & - & 34.51 & 0.8507 \\ 
			\rule[-1ex]{0pt}{3.5ex}
			WNNM \cite{WNNM} & Non-blind & - & 34.67 & 0.8646 \\
			\rule[-1ex]{0pt}{3.5ex}
			DnCNN+ \cite{DnCNN} & Non-blind & 668 K & 37.90 & 0.9430 \\
			\rule[-1ex]{0pt}{3.5ex}
			FFDNet+ \cite{FFDNet} & Non-blind & 825 K & 37.61 & 0.9415 \\
			\rule[-1ex]{0pt}{3.5ex}
			GCBD \cite{GCBD} & Blind & 561 K & 35.58 & 0.9217 \\
			\rule[-1ex]{0pt}{3.5ex}
			CBDNet \cite{CBDNet} & Blind & 4.4 M & 38.06 & 0.9421 \\
			\rule[-1ex]{0pt}{3.5ex}
			RIDNet \cite{RIDNet} & Blind & 1.5 M & 39.26 & 0.9528 \\
			\rule[-1ex]{0pt}{3.5ex}
			VDN \cite{VDN} & Blind & 7.8 M & 39.38 & 0.9518 \\
			\rule[-1ex]{0pt}{3.5ex}
			AINDNet(S) \cite{AINDNet} & Blind & 13.7 M & 39.53 & \textcolor{blue}{0.9561} \\
			\rule[-1ex]{0pt}{3.5ex}
			MIRNet \cite{MIRNet} & Blind & 31.8 M & \textcolor{red}{39.88} & \textcolor{red}{0.9563} \\
			\rule[-1ex]{0pt}{3.5ex}
			MPRNet \cite{zamir2021multi} & Blind & 15.7 M & \textcolor{blue}{39.80} & {0.9540} \\
			\rule[-1ex]{0pt}{3.5ex}
			InvDN \cite{InvDN} & Blind & 2.6 M & 39.57 & 0.9522 \\
			\hline
			\rule[-1ex]{0pt}{3.5ex}
			VDIR (Ours) &Blind & 2.2 M & 39.63 & 0.9528 \\
			\rule[-1ex]{0pt}{3.5ex}
			VDIR+ (Ours)&Blind & 2.2 M & {39.69} & 0.9532\\
			
			\Xhline{4\arrayrulewidth}			
		\end{tabular}
	}
\end{table}

\begin{table*}[!t]
	\caption{The average PSNR/SSIM results of blind super-resolution for scaling factor $\times 2$. The best results are highlighted in \textbf{bold}.}
	\begin{center}
		\resizebox{0.75\linewidth}{!}{
			\begin{tabular}{c|c|c|c|c|c|c|c}
				\Xhline{4\arrayrulewidth}
				\rule[-1ex]{0pt}{3.5ex}
				Kernel & Dataset & Bicubic & RCAN\cite{RCAN} & IKC\cite{IKC} & MZSR\cite{MZSR} & DIP-FKP \cite{FKP} + USRNet \cite{USRNet} & VDIR (Ours) \\
				\hline\hline
				
				\rule[-1ex]{0pt}{3.5ex}
				\multirow{3}{*}{$g_{2.0}^d$}&Set5& 28.73/0.8449 & 29.15/0.8601 & 29.05/0.8896 & 36.05/0.9439 & 30.65/0.8861 & \textbf{36.95/0.9506}\\
				\rule[-1ex]{0pt}{3.5ex}
				&BSD100 & 26.51/0.7157 & 26.89/0.7394 & 27.46/0.8156 & 31.09/0.8739 & 27.56/0.7484 & \textbf{31.57/0.8815}\\
				\rule[-1ex]{0pt}{3.5ex}
				&Urban100 & 23.70/0.7109 & 24.14/0.7384 & 25.17/0.8169 & 29.19/0.8838 & 25.96/0.7878 & \textbf{30.10/0.8987}\\
				\hline\hline
				
				\rule[-1ex]{0pt}{3.5ex}
				\multirow{3}{*}{$g_{ani}^d$}&Set5& 28.15/0.8265 & 28.42/0.8379 & 28.74/0.8565 & 34.78/0.9323 & 29.32/0.8694 &\textbf{36.25/0.9469}\\
				\rule[-1ex]{0pt}{3.5ex}
				&BSD100 & 26.00/0.6891 & 26.22/0.7062 & 26.44/0.7310 & 29.54/0.8297 & 26.78/0.7222 & \textbf{30.51/0.8594}\\
				\rule[-1ex]{0pt}{3.5ex}
				&Urban100 & 23.13/0.6796 & 23.35/0.6982 & 23.62/0.7239 & 27.34/0.8369 & 24.86/0.7517 & \textbf{28.58/0.8720}\\

				\Xhline{4\arrayrulewidth}
			\end{tabular}
		}
	\end{center}	
	\label{tab:SR}
\end{table*}

\begin{table*}[!t]
	\caption{The average PSNR/SSIM results of blind super-resolution for scaling factor $\times 4$. Isotropic Gaussian kernel $g_{2.0}^d$ is used for the blur kernel. The best results are highlighted in \textbf{bold}.}
	\begin{center}
		\resizebox{0.75\linewidth}{!}{
			\begin{tabular}{c|c|c|c|c|c|c}
				\Xhline{4\arrayrulewidth}
				\rule[-1ex]{0pt}{3.5ex}
				Dataset & Bicubic & RCAN\cite{RCAN} & IKC\cite{IKC} & MZSR\cite{MZSR} & DIP-FKP \cite{FKP} + USRNet \cite{USRNet}& VDIR (Ours) \\
				\hline\hline
				
				\rule[-1ex]{0pt}{3.5ex}
				Set5& 24.74/0.7321 & 23.92/0.7283 & 24.01/0.7322 & 30.50/0.8704& 28.14/0.8427 &\textbf{31.46/0.8884}\\
				\rule[-1ex]{0pt}{3.5ex}
				BSD100 & 24.01/0.5998 & 23.16/0.5918 & 23.12/0.5939 & 26.89/0.7168 & 25.64/0.6720 & \textbf{27.27/0.7271}\\
				\rule[-1ex]{0pt}{3.5ex}
				Urban100 & 21.16/0.5811 & 19.52/0.5400 & 19.81/0.5583 & 24.65/0.7394 & 23.62/0.7102 & \textbf{25.62/0.7730}\\
								
				\Xhline{4\arrayrulewidth}
			\end{tabular}
		}
	\end{center}	
	\label{tab:SR4}
\end{table*}

\subsection{Implementation Details}
For our VDIR, we set $N=5$ and $D=5$, which amounts to about $2.2 M$ parameters, including the restorer and the encoder. We adopt Adam optimizer with $\beta_1 =0.9$ and $\beta_2=0.999$ with initial learning rate $2 \times 10^{-4}$. The learning rate is decayed by half in every $100,000$ iterations, until it reaches $2\times 10^{-5}$.

\section{Image Denoising}
We perform two image denoising tasks: blind AWGN image denoising and real-noise denoising.

\subsection{Results on AWGN Denoising}
For the synthetic Gaussian noise, we train with DIV2K \cite{DIV2K} dataset which includes $800$ high-resolution images, and add synthetic Gaussian noise with noise level $\sigma \in [5, 70]$. We extract $96 \times 96$ patches and the batch size is set to $16$. We set $\beta=0.01$ and $\lambda_2=1$ for the hyperparameters, and $\sigma = f(T)$ to impose our prior knowledge, letting $\c$ to include the noise level information. The performance is evaluated with three color-image datasets: CBSD68\cite{BSD}, Kodak24, and Urban100\cite{Urban} with noise levels $\sigma = 10,~30,~50,~70$. We compare our method with several AWGN denoising methods: CBM3D\cite{BM3D}, DnCNN\cite{DnCNN}, FFDNet \cite{FFDNet}, UNLNet\cite{UNLNet}, and VDN\cite{VDN}. The results are presented in \tablename{~\ref{tab:AWGN}}.
The visualized results for the AWGN denoising and all the following experiments are provided in the {\em supplemental material}.

Note that CBM3D \cite{BM3D} and FFDNet \cite{FFDNet} are non-blind methods, whereas the rest are blind ones. In most cases, our VDIR shows the best results but requires a smaller number of parameters (2.2 M) compared to the second-best VDN \cite{VDN} (7.8 M). In conclusion, the results show that our VDIR surpasses other methods considering the tradeoff between the performance and the network capacity (the number of parameters).

\subsection{Results on Real-Noise Denoising}
For the real-world camera noise, we train the network using two datasets: Smartphone Image Denoising Dataset (SIDD) \cite{SIDD} and DIV2K \cite{DIV2K}, along with realistic synthetic noise \cite{CBDNet}.
SIDD is a collection of pairs of noisy and clean images from five smartphone cameras. It consists of $320$ image pairs for training.
For synthetic noise, we adopt the noise synthesis process of CBDNet \cite{CBDNet}, where the camera pipeline, including demosaicking and camera response functions, and heteroscedastic Gaussian noise are considered to mimic real-noise as much as possible.
We extract $256\times 256$ patches with the batch size $4$ for training. We set $\beta=0.01$ and $\lambda_2=0$, since we do not have the degradation information for the real-noise case.
We compare with several image denoisnig methods: BM3D \cite{BM3D}, WNNM \cite{WNNM}, DnCNN\cite{DnCNN}, TNRD\cite{TNRD}, FFDNet \cite{FFDNet}, GCBD \cite{GCBD}, CBDNet \cite{CBDNet}, RIDNet \cite{RIDNet}, VDN \cite{VDN}, AINDNet \cite{AINDNet}, MIRNet \cite{MIRNet}, MPRNet~\cite{zamir2021multi}, and InvDN \cite{InvDN}.
For the evaluation, we use two widely used real-world image denoising benchmarks.

\begin{itemize}
	\item \textbf{SIDD:}
SIDD provides $1,280$ small patches for validation and $1,280$ for test, which are visually similar to training images. Ground-truth patches for the validation set are provided but not for the test set.
\item \textbf{DND:}
Darmstadt Noise Dataset (DND) consists of $50$ images with real-noise from $50$ scenes from four consumer cameras. Then, the images are further cropped by the provider, which results in $1,000$ small patches with a size of $512 \times 512$.

Both test sets do not provide the ground-truth image; therefore, the evaluation can only be performed through online submission.
\end{itemize}

Overall results on two benchmarks are listed in \tablename{~\ref{tab:SIDD}} and \ref{tab:DND}. It is observed that our method shows comparable results to others, on both benchmarks in terms of PSNR and SSIM \cite{SSIM}. Note that we also demonstrate the results with self-ensemble \cite{Sevenways}, which is denoted with `+' sign.

Compared to conventional methods such as BM3D \cite{BM3D} and WNNM \cite{WNNM}, it is evident that the data-driven methods are superior in real-noise denoising due to the power of CNN and large-scale external dataset. Also, many assumptions and constraints in the conventional methods might have resulted in deteriorated performance. In the case of CNN-based methods for AWGN such as DnCNN \cite{DnCNN} and FFDNet \cite{FFDNet}, they suffer from domain shift between i.i.d. Gaussian and real-noise. On the other hand, most of the recent CNN-based real image denoising methods \cite{CBDNet, RIDNet, VDN, AINDNet, InvDN} show competitive results, achieving state-of-the-art performances.

For a fair comparison, we also denote the number of parameters for the CNN-based methods.
Note that MIRNet \cite{MIRNet} achieves PSNR gain against our method about 0.2 dB in DND, but the number of parameters is 31.8 M which exceeds an order of magnitude compared to ours. Also, the MPRNet \cite{zamir2021multi} achieves higher performance (0.17 dB) using a larger network (15.7 M parameters).
Since our method solves simpler sub-problems conditioned on the latent variable $\c$, it requires a smaller network than other denoising networks, which are the na\"{i}ve models. In other words, the problem given to other methods is more complicated due to the difficulty of ill-posed real image denoising with diverse conditions. Therefore, our method achieves state-of-the-art performances in real-world image denoising with fewer network parameters than others.

\begin{table*}[!t]
	\caption{The average PSNR/SSIM results on JPEG compression artifacts reduction. The best results are highlighted in \textbf{bold}. The numbers for compared methods are from the original papers, and the blank cells are not reported cases.}
	\begin{center}
		\resizebox{0.98\linewidth}{!}{
			\begin{tabular}{c|c|c|c|c|c|c|c}
				\Xhline{4\arrayrulewidth}
				\rule[-1ex]{0pt}{3.5ex}
				QF & Dataset & JPEG & AR-CNN \cite{ARCNN} & Guo \emph{et al.} \cite{One-to-many} & MemNet \cite{MemNet} & Galteri \emph{et al.} \cite{GACAR} & VDIR (Ours)\\
				\hline\hline
				
				\rule[-1ex]{0pt}{3.5ex}
				\multirow{3}{*}{$10$}&Classic5&27.82/0.7800& 29.03/0.8108 & 29.62/0.8260&29.68/0.8282&-& \textbf{29.91/0.8333}\\
				\rule[-1ex]{0pt}{3.5ex}
				&LIVE1  & 27.77/0.7905 & 28.96/0.8217 & 29.36/0.8300&29.47/0.8337&29.45/0.8340& \textbf{29.56/0.8348}\\
				\rule[-1ex]{0pt}{3.5ex}
				&Urban100& 26.33/0.8099 & 28.06/0.8515 & 29.01/0.9710& 29.14/0.8741& - &\textbf{29.59/0.8803}\\
				\hline
				
				\rule[-1ex]{0pt}{3.5ex}
				\multirow{3}{*}{$20$}&Classic5&30.12/0.8541&31.15/0.8691 & 31.77/0.8790&31.90/0.8658&-& \textbf{32.12/0.8850}\\
				\rule[-1ex]{0pt}{3.5ex}
				&LIVE1  & 30.07/0.8683 & 31.29/0.8871 & 31.68/0.8950&31.83/0.8973&31.77/0.8960&\textbf{31.95/0.8989}\\
				\rule[-1ex]{0pt}{3.5ex}
				&Urban100& 28.57/0.8761 & 30.29/0.9024&31.39/0.9170&31.61/0.9206&-&\textbf{32.17/0.9257}\\
				\hline
				
				\rule[-1ex]{0pt}{3.5ex}
				\multirow{3}{*}{$30$}&Classic5&31.48/0.8844& 32.51/0.8963&33.04/0.9030&32.90/0.8988&-&\textbf{33.38/0.9066}\\
				\rule[-1ex]{0pt}{3.5ex}
				&LIVE1  & 31.41/0.9000 & 32.67/0.9161&33.09/0.9210&32.94/0.9169&33.15/0.9220&\textbf{33.38/0.9245}\\
				\rule[-1ex]{0pt}{3.5ex}
				&Urban100& 30.00/0.9052 & 31.94/0.9279&32.90/0.9380&32.74/0.9348&-&\textbf{33.70/0.9437}\\
				\hline
				
				\rule[-1ex]{0pt}{3.5ex}
				\multirow{3}{*}{$40$}&Classic5&32.43/0.9011 & 33.32/0.9098&33.89/0.9150&33.36/0.9060&-&\textbf{34.23/0.9187}\\
				\rule[-1ex]{0pt}{3.5ex}
				&LIVE1  & 32.35/0.9173 & 33.61/0.9303&34.09/0.9360&33.48/0.9248&34.09/0.9350&\textbf{34.37/0.9381}\\
				\rule[-1ex]{0pt}{3.5ex}
				&Urban100& 31.07/0.9220 & 32.80/0.9391&33.98/0.9490&33.22/0.9405&-&\textbf{34.73/0.9535}\\
				\hline\hline
				
				\rule[-1ex]{0pt}{3.5ex}
				\multirow{3}{*}{$70$}&Classic5&34.98/0.9344 & - & - &-&-&\textbf{36.40/0.9431}\\
				\rule[-1ex]{0pt}{3.5ex}
				&LIVE1  & 35.13/0.9512 & - &-&-&-&\textbf{37.15/0.9638}\\
				\rule[-1ex]{0pt}{3.5ex}
				&Urban100& 34.32/0.9563 & -&-&-&-&\textbf{37.51/0.9723}\\
				\hline
				
				\rule[-1ex]{0pt}{3.5ex}
				\multirow{3}{*}{$80$}&Classic5&36.44/0.9478 & - & - &-&-&\textbf{37.63/0.9537}\\
				\rule[-1ex]{0pt}{3.5ex}
				&LIVE1  & 36.87/0.9641 & -&-&-&-&\textbf{38.81/0.9733}\\
				\rule[-1ex]{0pt}{3.5ex}
				&Urban100& 36.69/0.9697 & -&-&-&-&\textbf{39.16/0.9796}\\
				
				\Xhline{4\arrayrulewidth}
			\end{tabular}
		}
	\end{center}	
	\label{tab:JPEG}
\end{table*}

\begin{table}[!t]
	\caption{The average PSNR results of JPEG compression artifacts reduction on sRGB color images.}
	\begin{center}
		\resizebox{0.98\linewidth}{!}{
			\begin{tabular}{c|c|c|c|c|c|c|c|c|c}
				\Xhline{4\arrayrulewidth}
				\rule[-1ex]{0pt}{3.5ex}
				Dataset & QF & 10 & 20 & 30 & 40 & 50 & 60 & 70 & 80 \\
				\hline\hline
				
				\rule[-1ex]{0pt}{3.5ex}
				\multirow{3}{*}{LIVE1}&JPEG &25.69&28.06&29.37&30.28&31.03&31.77&32.77&34.23\\
				\rule[-1ex]{0pt}{3.5ex}
				&VDIR & 27.57&30.03&31.38&32.32&33.08&33.81&34.79&36.17\\
				\cline{2-10}
				\rule[-1ex]{0pt}{3.5ex}
				&$\Delta$PSNR& 1.88&1.97&2.01&2.04&2.05&2.04&2.02&1.94\\
				\hline\hline
				
				\rule[-1ex]{0pt}{3.5ex}
				\multirow{3}{*}{Urban100}&JPEG&24.47&26.63&27.96&28.93&29.70&30.48&31.67&33.56\\
				\rule[-1ex]{0pt}{3.5ex}
				&VDIR & 27.34&29.72&31.08&31.97&32.64&33.33&34.26&35.47\\
				\cline{2-10}
				\rule[-1ex]{0pt}{3.5ex}
				&$\Delta$PSNR&2.87&3.09&3.12&3.04&2.94&2.85&2.59&1.91\\
				
				\Xhline{4\arrayrulewidth}
			\end{tabular}
		}
	\end{center}	
	\label{tab:JPEG_Color}
\end{table}

\begin{figure}	
	\captionsetup{justification=centering}
	\begin{center}
		\begin{subfigure}[t]{0.49\linewidth}
			\centering
			\includegraphics[width=1\columnwidth]{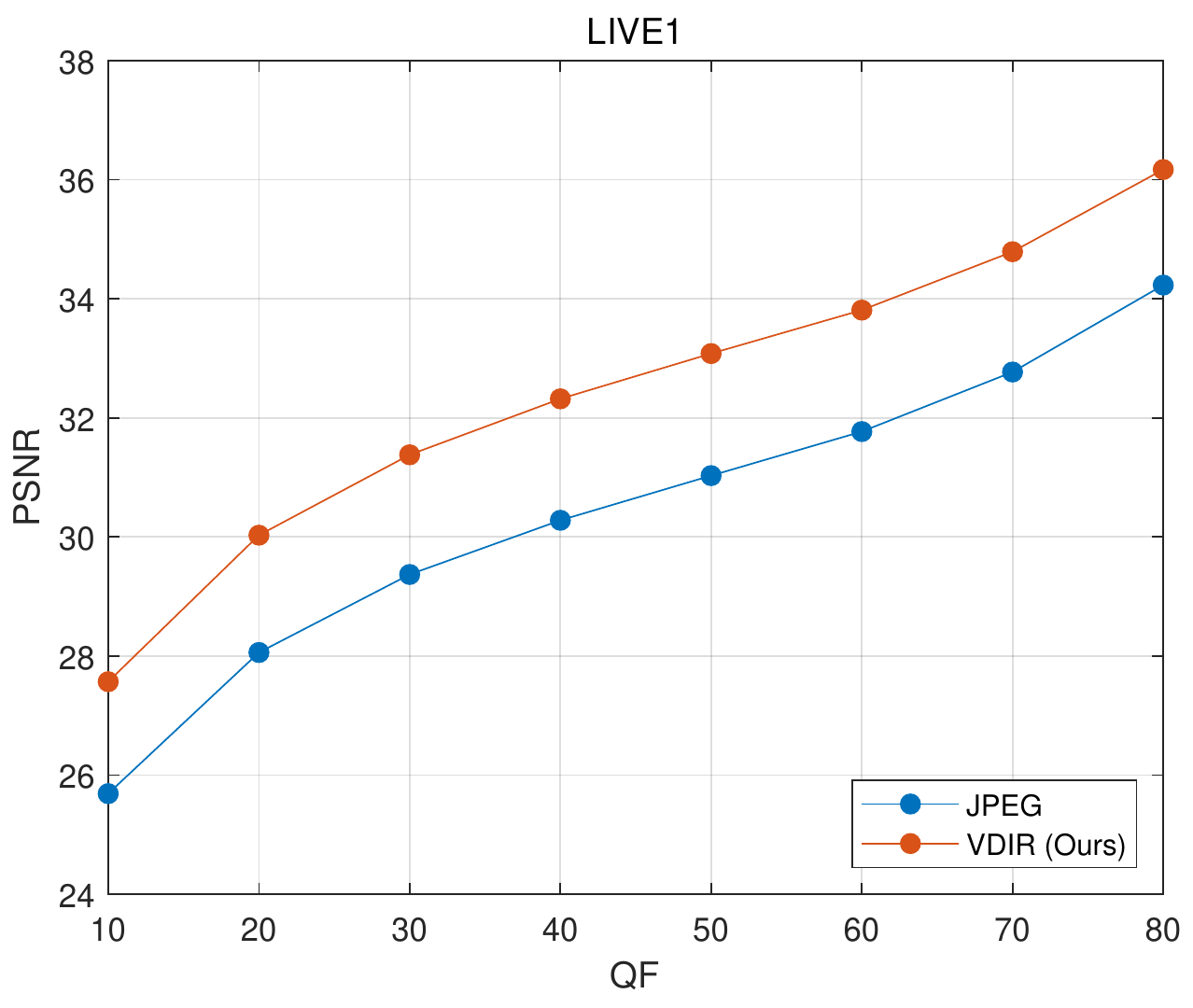}
			\caption{LIVE1 \cite{LIVE1}}
		\end{subfigure}			
		\begin{subfigure}[t]{0.49\linewidth}
			\centering
			\includegraphics[width=1\columnwidth]{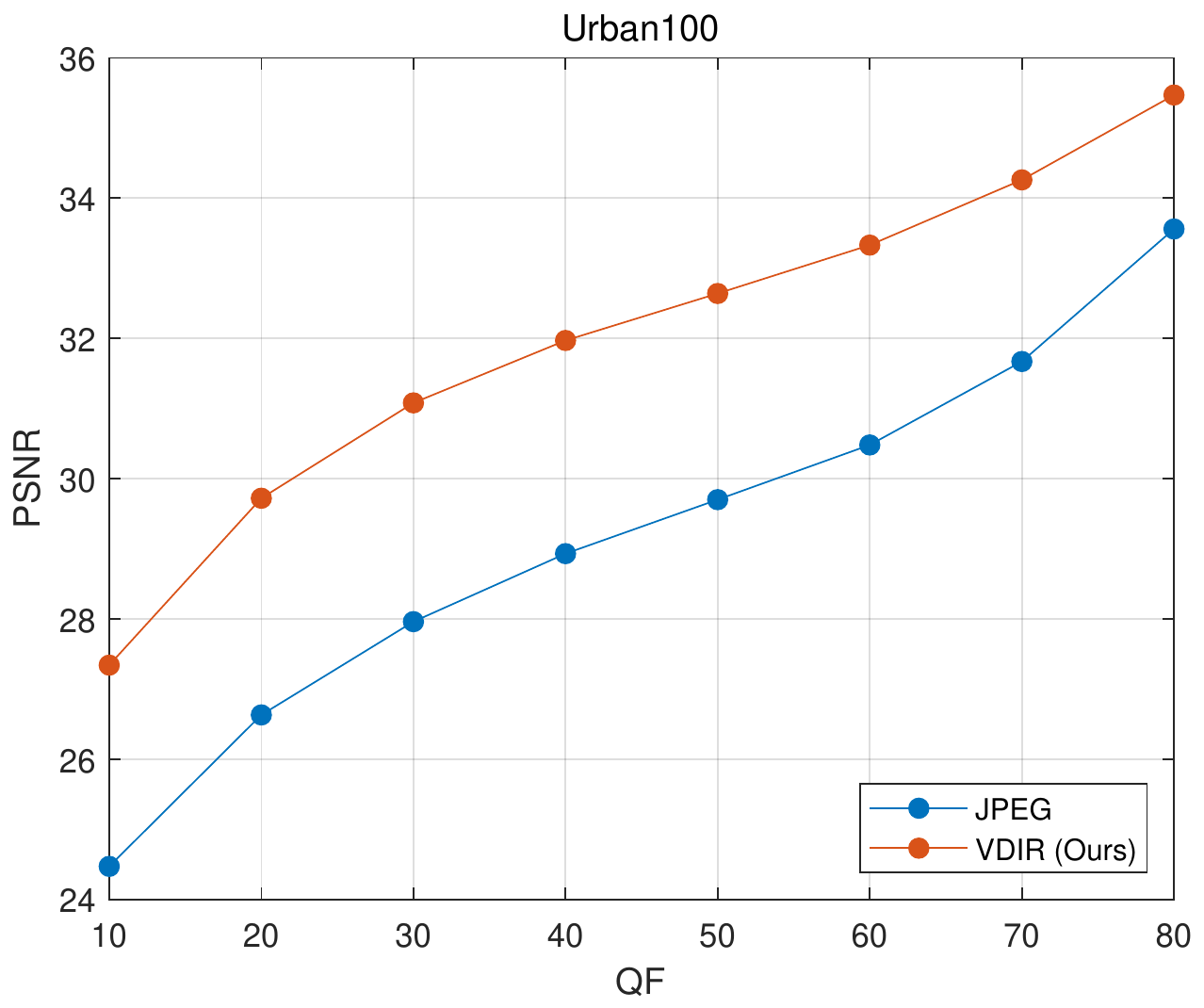}
			\caption{Urban100 \cite{Urban}}
		\end{subfigure}			
	\end{center}
	\caption{PSNR vs. QF curves of artifacts reduction results on two datasets.}
	\label{fig:JPEG_Color}
\end{figure}

\section{Blind Image Super-Resolution}
To verify the generality of our method, we also evaluate our method on blind image super-resolution task following the settings of \cite{MZSR}.
We train with DIV2K \cite{DIV2K} dataset with various Gaussian kernels for blur kernels $k \in \mathbb{R}^{15\times 15}$ with the same setting of MZSR \cite{MZSR}. We adopt direct decimation for sub-sampling method. Concretely, we extract $48 \times 48$ as our low-resolution (LR) patches with the batch size $16$ and trained two networks for each scaling factor: $\times2$ and $\times4$. We set $\beta=0.01$, $\lambda_2=1$, and $f(T) = k$ for the loss function. The evaluations are done in Y (luminance) channel of YCbCr colorspace, and two Gaussian kernels (isotropic and anisotropic) are considered for $\times 2$.
Specifically, $g_{2.0}^d$ denotes an isotropic Gaussian kernel with width $2.0$ followed by direct subsampling. $g_{ani}^d$ denotes an anisotropic Gaussian kernel with widths $4.0$ and $1.0$ and rotated with $-0.5$ radian.
The overall results are presented in \tablename{~\ref{tab:SR}} and \tablename{~\ref{tab:SR4}.

We compare with several recent super-resolution methods: RCAN \cite{RCAN}, IKC \cite{IKC}, MZSR \cite{MZSR}, and DIP-FKP \cite{FKP} with USRNet \cite{USRNet}.
As shown in the tables, RCAN \cite{RCAN}, which is a large model trained only for the bicubic downsampling scenario, suffers from domain shift, delivering inferior results to others. Blind super-resolution method such as IKC \cite{IKC} shows improved results but still not plausible. MZSR \cite{MZSR}, which adopts a meta-learning strategy for fast adaptation, shows competitive results.
Another blind super-resolution method, flow-based kernel prior \cite{FKP} combined with USRNet \cite{USRNet} dubbed ``DIP-FKP + USRNet'' achieves plausible results. However, it is noteworthy that a thousand iterations are required to obtain a super-resolved image.
Among all comparisons, our method outperforms others in the blind super-resolution task in terms of PSNR and SSIM.

\section{JPEG Compression Artifacts Reduction}

We choose dominantly used JPEG \cite{JPEG} for the experiments on compression artifacts reduction. Specifically, we evaluate our method to reduce artifacts in images compressed by JPEG with a wide range of quality factors (QFs) in $[10, 80]$.
We train with DIV2K \cite{DIV2K} dataset with JPEG compression of MATLAB with QFs in $[10, 80]$ with $10$ intervals.
In particular, we extract $96 \times 96$ patches with the batch size $16$ and we set $\beta=0.01$, $\lambda_2=1$, and $f(T) = QF$ for the overall loss. We present two models: Y-model and RGB-model. Particularly, following the convention of previous works and for a fair comparison, we present a single channel model, which is for the Y channel of YCbCr colorspace. For more practical usage, we present a color model which is trained on sRGB images.

The performance is evaluated with three datasets: Classic5, LIVE1\cite{LIVE1}, and Urban100\cite{Urban} with $QF = 10,~20,~30,~40,~70,~80$. We compare our method with other methods: AR-CNN\cite{ARCNN}, Guo \emph{et al.}\cite{One-to-many}, MemNet\cite{MemNet}, and Galteri \emph{et al.} \cite{GACAR}. The results are shown in \tablename{~\ref{tab:JPEG}}.

Most previous methods are targeting highly compressed images (low quality factors), and mostly are non-blind methods except MemNet\cite{MemNet}. It is observed from the table that our method can deal with a wide range of quality factors in a blind manner, and also it outperforms others with comparable model sizes. Importantly, many images we face every day online are JPEG, and their compression rates are very high (high QFs), but they still have undesirable artifacts. Previous methods did not deal with such images, for example, the quality factor above $70$, but our method shows consistent PSNR/SSIM gain even on high quality factors of $70$ and $80$.
For more practical application, we also present our color model results in \tablename{~\ref{tab:JPEG_Color}} and the performance curves in \figurename{~\ref{fig:JPEG_Color}}. Our VDIR shows consistent gains in the wide range of QFs with both datasets. 

\section{Analysis}

\subsection{Analysis on the Latent Variable}
\begin{figure*}
	\begin{center}		
		\captionsetup{justification=centering}
		\begin{subfigure}[t]{0.19\linewidth}
			\centering
			\includegraphics[width=1\columnwidth]{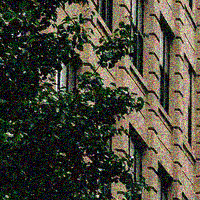}
			\caption{Noisy image $\sigma=30$ \\ 19.72 dB}
		\end{subfigure}
		\begin{subfigure}[t]{0.19\linewidth}
			\centering
			\includegraphics[width=1\columnwidth]{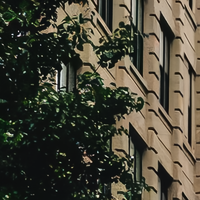}
			\caption{The same input \\ 27.22 dB}
			\label{fig:latent1b}
		\end{subfigure}
		\begin{subfigure}[t]{0.19\linewidth}
			\centering
			\includegraphics[width=1\columnwidth]{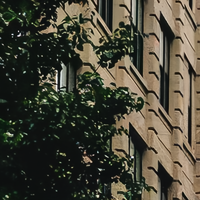}
			\caption{Flipped input \\ 27.10 dB}
			\label{fig:latent1c}
		\end{subfigure}
		\begin{subfigure}[t]{0.19\linewidth}
			\centering
			\includegraphics[width=1\columnwidth]{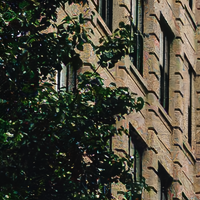}
			\caption{Input with $\sigma=10$ \\ 25.96 dB}
			\label{fig:latent1d}
		\end{subfigure}
		\begin{subfigure}[t]{0.19\linewidth}
			\centering
			\includegraphics[width=1\columnwidth]{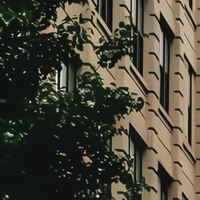}
			\caption{Input with $\sigma=50$ \\ 26.18 dB}
			\label{fig:latent1e}
		\end{subfigure}
	\end{center}
	\caption{Latent space manipulation by feeding different input images to the encoder. For the denoiser, (a) we feed an input image with $\sigma=30$. For the encoder, we feed four different inputs: (b) the same noisy image, (c) flipped noisy image, and (d-e) the same image with different noise levels.}
	\label{fig:latent1}
\end{figure*}

\begin{figure*}
	\begin{center}		
		\captionsetup{justification=centering}
		\begin{subfigure}[t]{0.19\linewidth}
			\centering
			\includegraphics[width=1\columnwidth]{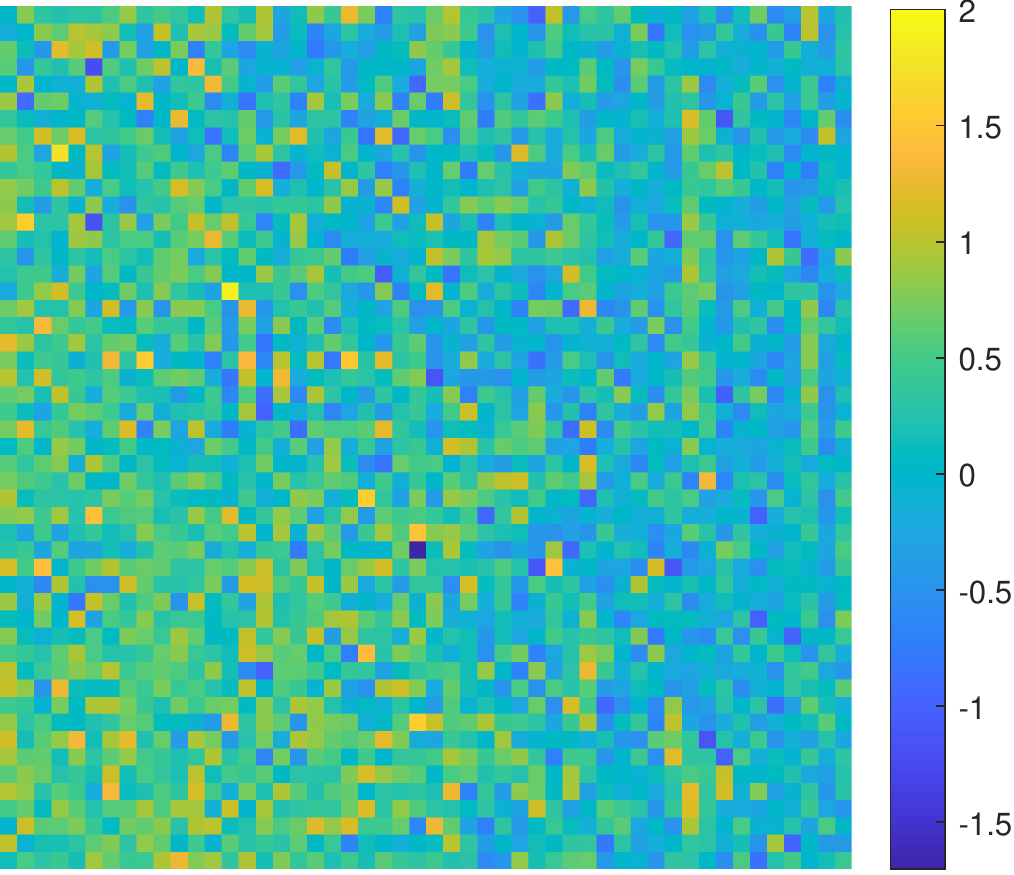}
			\caption{The true noisy input ($\sigma=30$)}
			\label{fig:con1a}
		\end{subfigure}
		\begin{subfigure}[t]{0.19\linewidth}
			\centering
			\includegraphics[width=1\columnwidth]{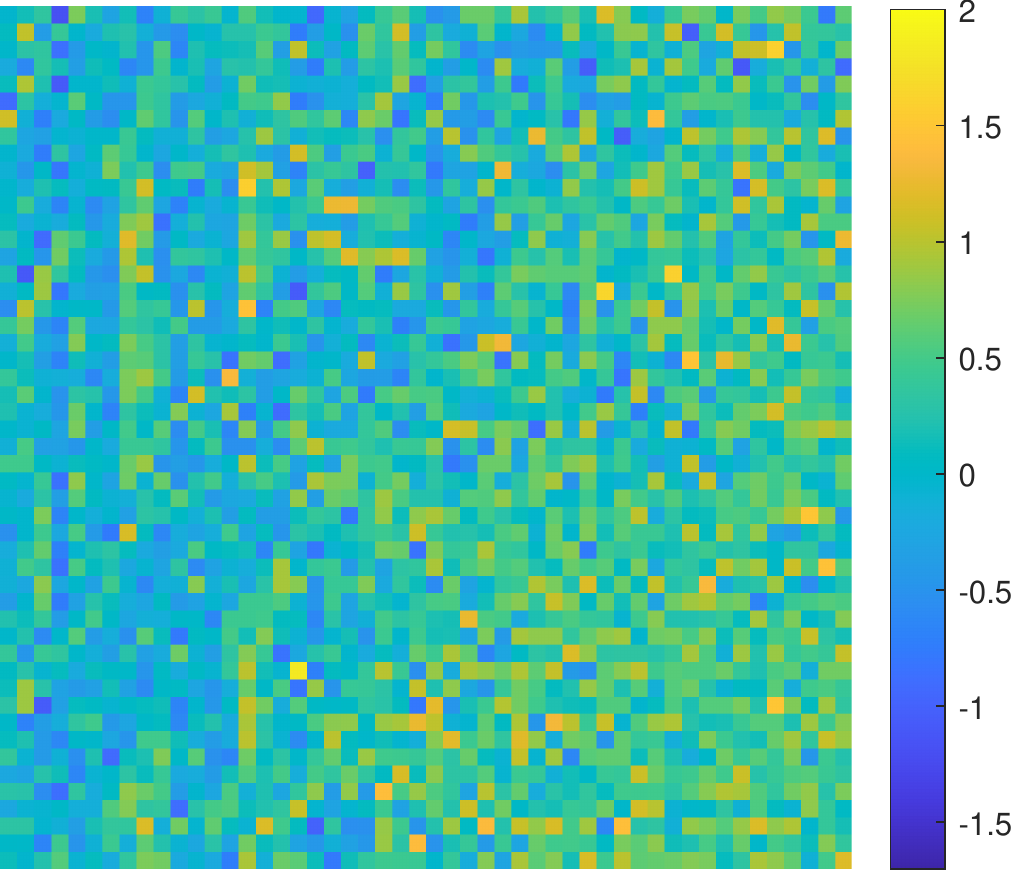}
			\caption{Flipped input}
			\label{fig:cond1b}
		\end{subfigure}
		\begin{subfigure}[t]{0.19\linewidth}
			\centering
			\includegraphics[width=1\columnwidth]{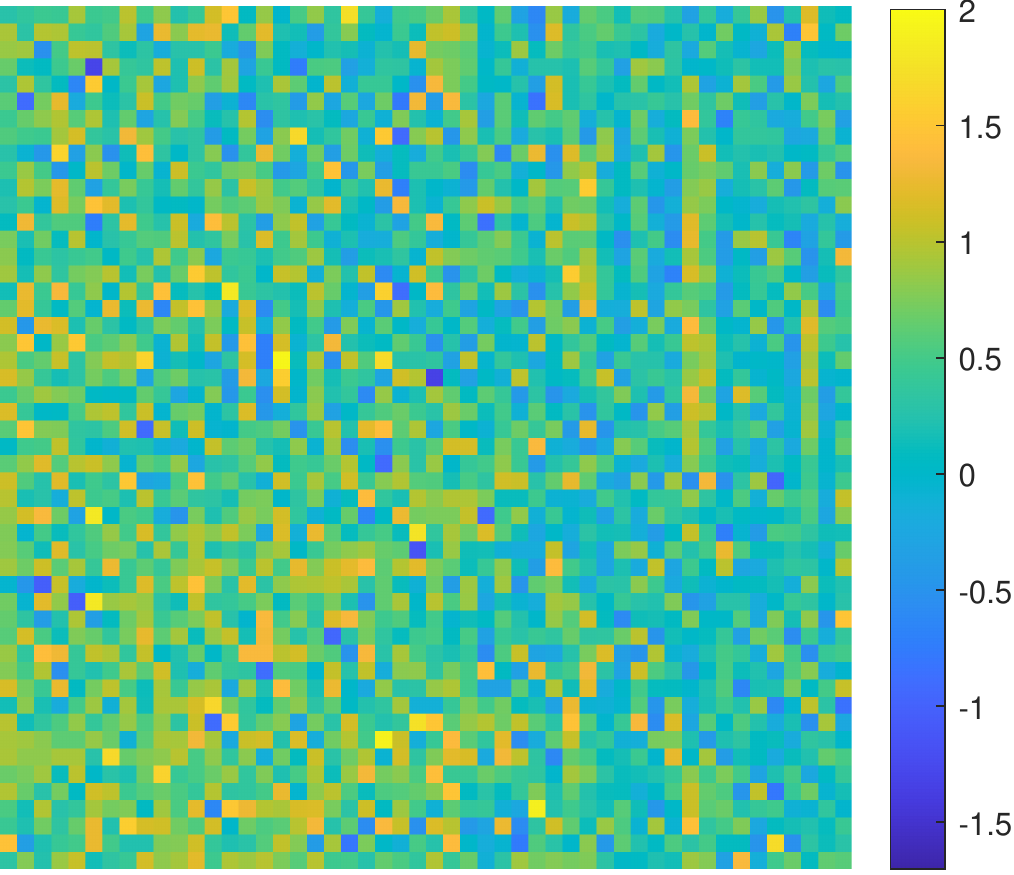}
			\caption{Input with $\sigma=10$}
			\label{fig:cond1c}
		\end{subfigure}
		\begin{subfigure}[t]{0.19\linewidth}
			\centering
			\includegraphics[width=1\columnwidth]{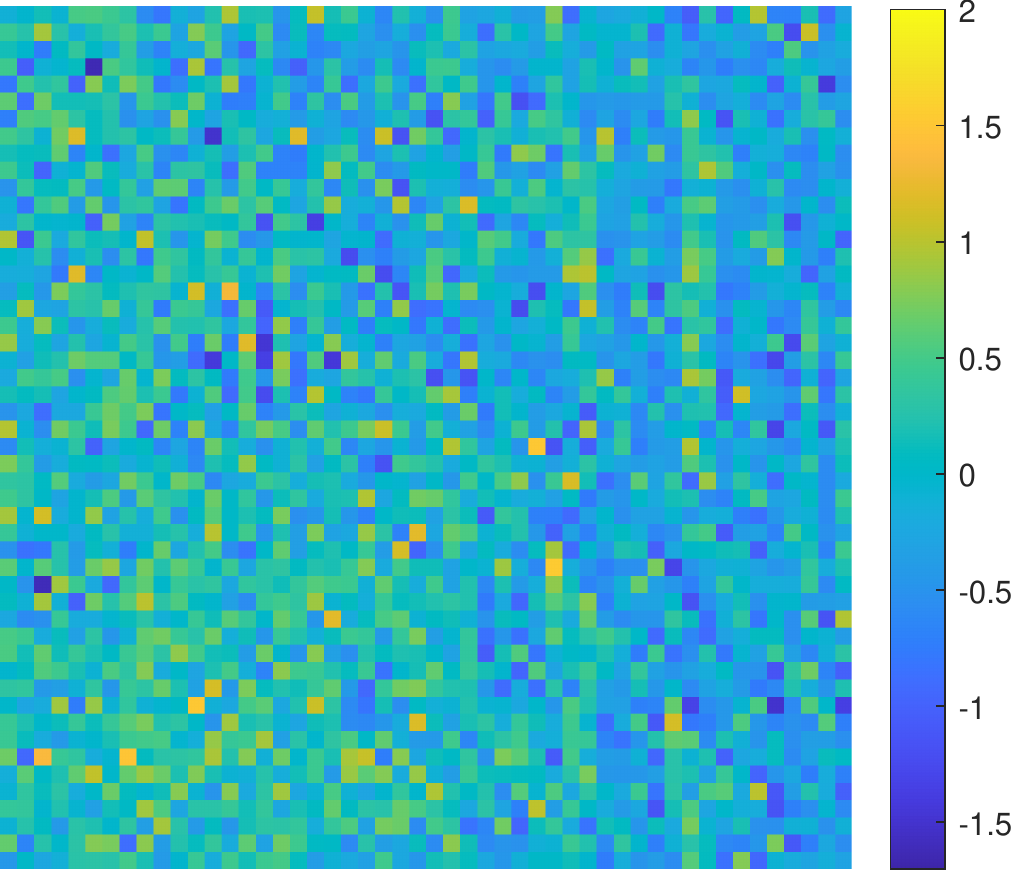}
			\caption{Input with $\sigma=50$}
			\label{fig:cond1d}
		\end{subfigure}

	\end{center}
	\caption{Latent space visualization of \figurename{~\ref{fig:latent1}}. It can be seen that the latent variable reflects the noise level and image contents very well.}
	\label{fig:cond1}
\end{figure*}

\begin{figure}
	\begin{center}
		\begin{subfigure}[t]{0.49\linewidth}
			\centering
			\includegraphics[width=1\columnwidth]{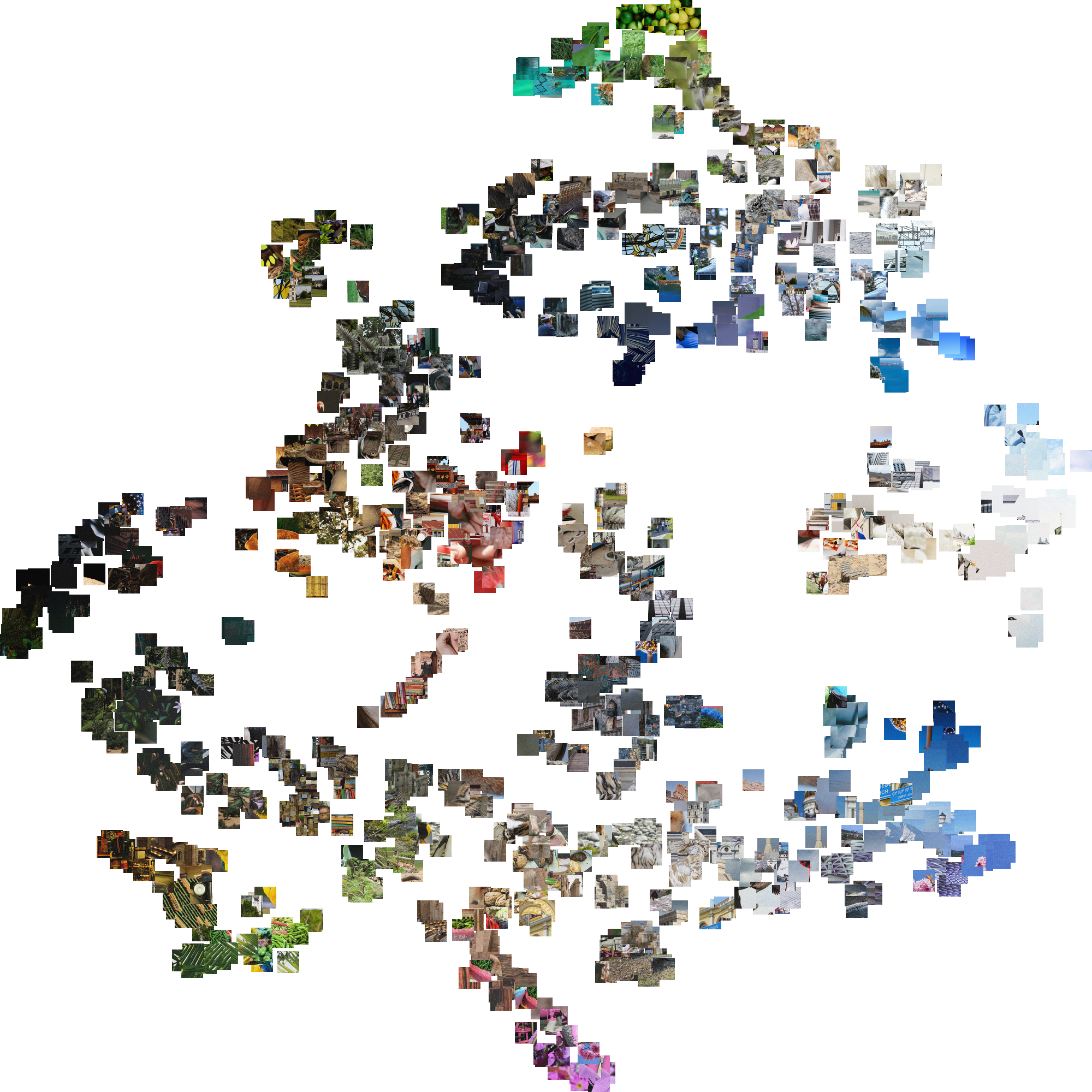}
			\caption{Patches}
			\label{fig:tSNE_image}
		\end{subfigure}
		\begin{subfigure}[t]{0.49\linewidth}
			\centering
			\includegraphics[width=1\columnwidth]{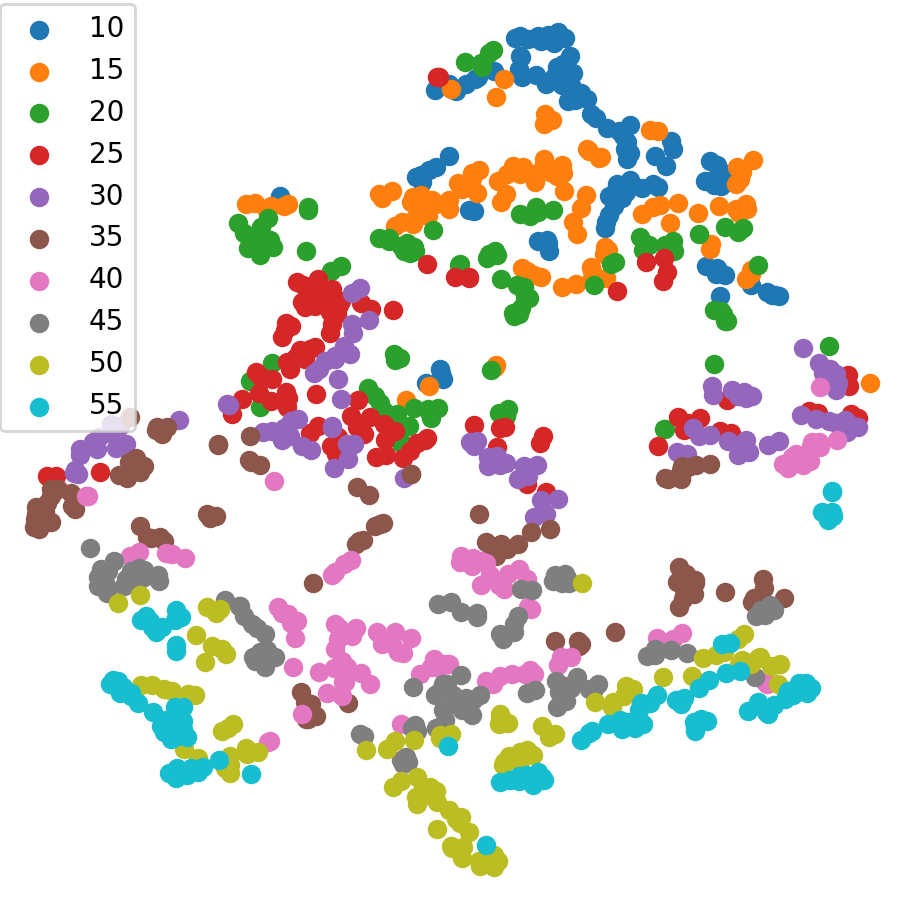}
			\caption{Noise levels}
			\label{fig:tSNE_noise}
		\end{subfigure}
	\end{center}
	\caption{t-SNE visualizations on Gaussian noise. (a) t-SNE visualization with patches. (b) Corresponding noise levels presented with different colors.}
	\label{fig:tSNE}
\end{figure}

\begin{figure}
	\begin{center}
		\begin{subfigure}[t]{0.49\linewidth}
			\centering
			\includegraphics[width=1\columnwidth]{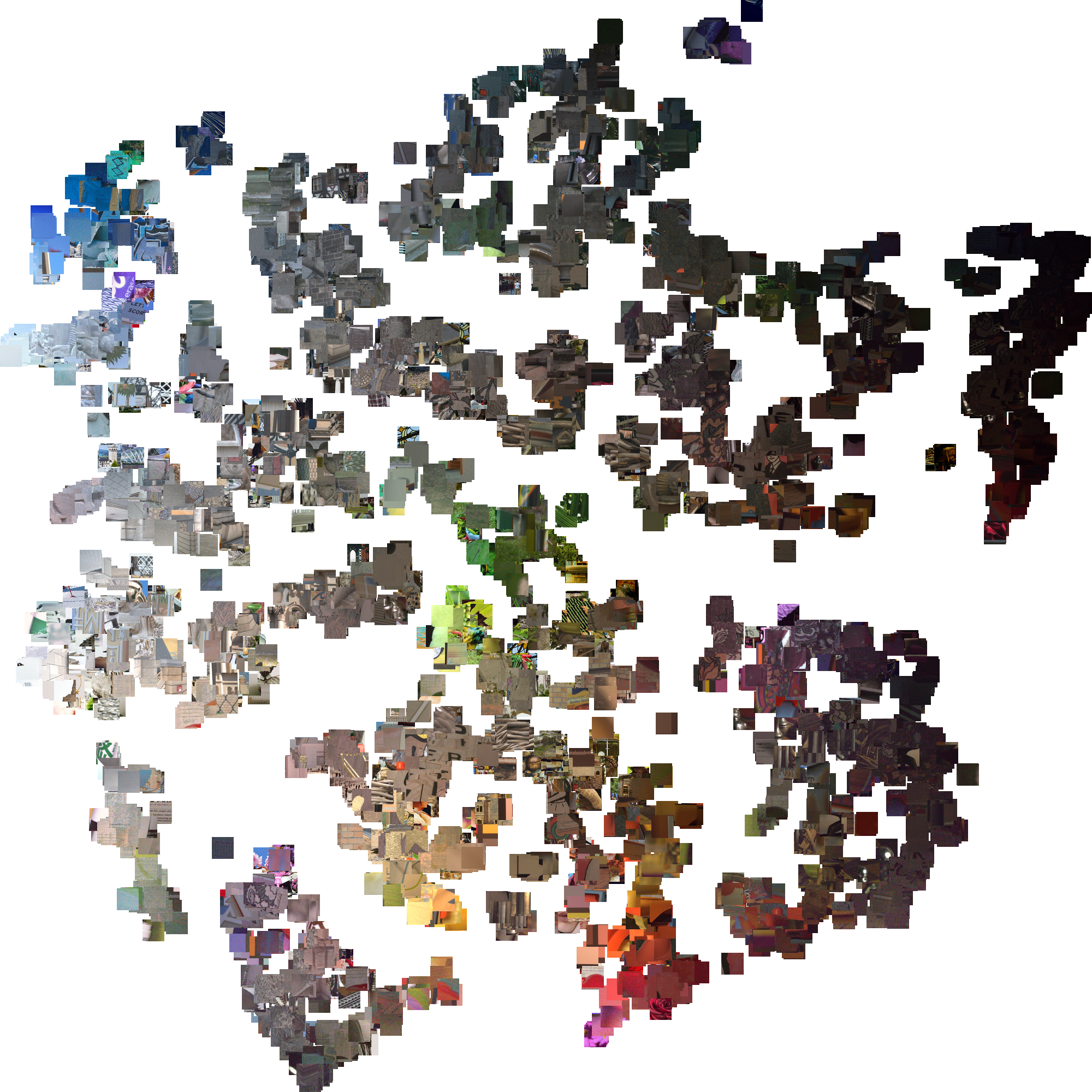}
			\caption{Patches}
			\label{fig:tSNE2_image}
		\end{subfigure}
		\begin{subfigure}[t]{0.49\linewidth}
			\centering
			\includegraphics[width=1\columnwidth]{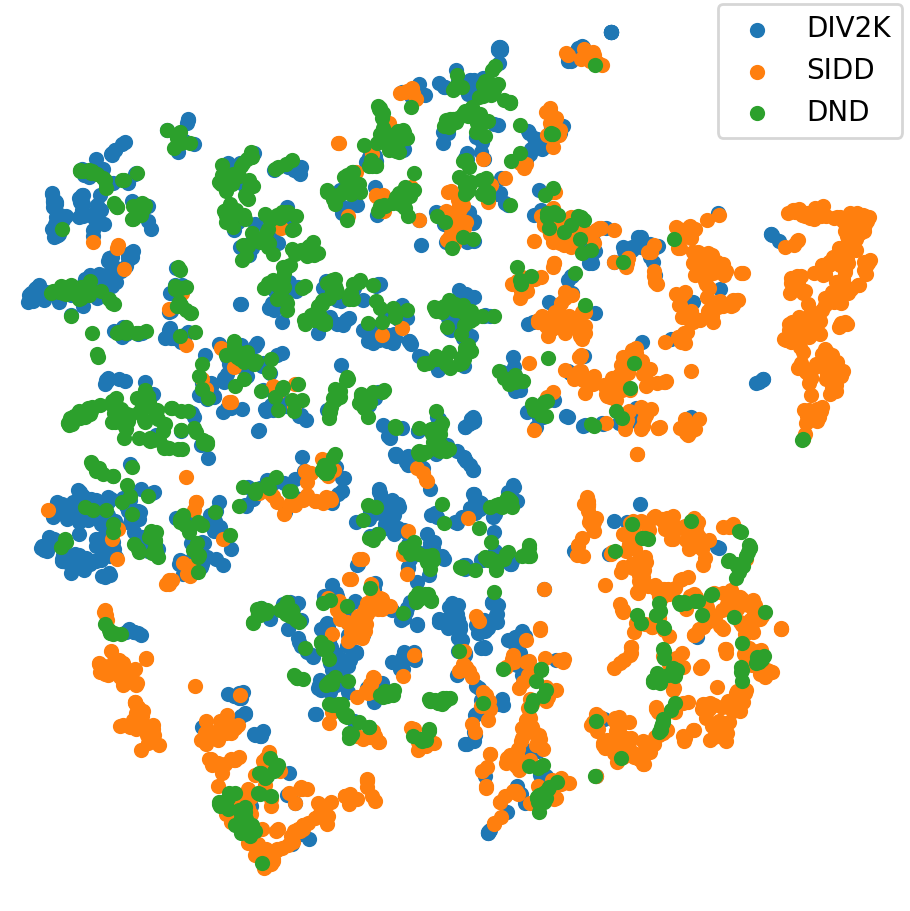}
			\caption{Datasets}
			\label{fig:tSNE2_data}
		\end{subfigure}
	\end{center}
	\caption{t-SNE visualizations on real-noise. (a) t-SNE visualization with patches. (b) Corresponding datasets.}
	\label{fig:tSNE2}
\end{figure}



In this subsection, we look into the latent space to investigate what kind of information is embedded.

\subsubsection{Latent Variable Manipulation}
First, to investigate the role of the latent variable $\c$, we manipulate the latent variable by feeding different inputs to the encoder of our AWGN denoiser. We feed the same noisy image with $\sigma = 30$ to the denoiser and evaluate how the information encoded in $\c$ affects the output image.
The results are presented in \figurename{~\ref{fig:latent1}}. As the latent variable contains the information of the noisy image manifold, it provides an additional prior to the denoiser. As expected, the best result is obtained when $\c$ is from the same noisy image (\figurename{~\ref{fig:latent1b}}). When $\c$ is from the same image but with a lower noise level (\figurename{~\ref{fig:latent1d}}), the output still has the noise patterns because $\c$ guides the noise level information. On the contrary, when $\c$ is from the same image with a higher noise level (\figurename{~\ref{fig:latent1e}}), the resulting image is overly smoothed.

Additionally, we feed a flipped image of the noisy image to obtain $\c$, which has the same noise level but different contents (\figurename{~\ref{fig:latent1c}}). Notably, the PSNR drop is not as significant as the other cases, but some artifacts and degradation exist due to the difference in semantics. The noise is left in the building part, whereas the tree part is over-smoothed. Commonly, trees have more textures than buildings. The latent variable reflects such common knowledge and works as additional prior information that guides the denoiser and how much the noise should be removed.

We visualize the latent variable of each case in \figurename{~\ref{fig:cond1}}. It is observed that the latent variable reflects the content information along with the noise level information. Specifically, the values vary in accordance with the noise level, and also are different between the building and tree regions. 

\subsubsection{t-SNE Visualization}
Secondly, we demonstrate t-SNE \cite{tSNE} visualization of $\c$ in \figurename{~\ref{fig:tSNE}}. In particular, $\c$ is average-pooled to generate global abstract of a patch as $AvgPool(\c) \in \mathbb{R}^n$. We present results on $1,000$ patches which are randomly extracted from DIV2K validation set, and the Gaussian noises are added with noise levels $10$ to $55$ at $5$ intervals.

As shown in \figurename{~\ref{fig:tSNE_image}}, the latent embedding well represents the content information, generating clusters of similar-looking patches. Also, the noise information is well embedded as shown in \figurename{~\ref{fig:tSNE_noise}}, where the patches with similar noise levels are closely located, and the embedding varies continuously from low $\sigma$ to high $\sigma$. In conclusion, the latent space, which is suitable for denoising tasks, contains not only noise level information but also global content information.

Moreover, we also present t-SNE visualization of the real-noise denoising in \figurename{~\ref{fig:tSNE2}}. We extract patches from three datasets: SIDD \cite{SIDD} validation set, DND \cite{DND}, and DIV2K validation set \cite{DIV2K} with synthetic noise following \cite{CBDNet}.
As shown in \figurename{~\ref{fig:tSNE2_image}}, the latent embedding is highly correlated to the content information such as colors and intensities. It might be connected to the common knowledge that the noise from the real world is signal-dependent. Interestingly, based on \figurename{~\ref{fig:tSNE2_data}}, the latent code captures different characteristics between the noise from SIDD\cite{SIDD} and DND\cite{DND}, despite we did not inject any supervision about the dataset. Concretely, they are separately clustered, and we might infer that there exists a domain gap between them.
In other words, our latent code sees the difference in the noise distribution of smartphone cameras and commercial cameras. Rather, the synthetic noise based on \cite{CBDNet} better mimic the characteristics of the noise from DND compared to SIDD \cite{SIDD} based on our observation.

To verify this observation, we train a baseline network which is the restorer without the encoder-decoder network with two different training datasets: DIV2K \cite{DIV2K} with synthetic noise \cite{CBDNet}, and SIDD \cite{SIDD}. The ablation results are presented in \tablename{~\ref{tab:dataset}}. Interestingly, training with synthetic datasets shows poor results in SIDD, whereas the results on DND are quite moderate. In contrast, training with the SIDD training dataset shows decent results both in the SIDD validation set and DND. The overall observations are quite consistent with our latent embeddings, which are learned in an unsupervised manner. We may conclude that our latent representation provides informative cues relevant to the denoising objective.

\begin{table}[!t]
	\caption{Ablation study between two datasets: DND and SIDD}
	\begin{center}
			\begin{tabular}{cc|cc}
				\hline
				\multicolumn{2}{c|}{\rule[-1ex]{0pt}{3.5ex}\multirow{2}{*}{Dataset}} & \multicolumn{2}{c}{Training} \\
				\rule[-1ex]{0pt}{3.5ex}
				&& DIV2K & SIDD \\
				\hline
				\rule[-1ex]{0pt}{4.0ex}
				{\multirow{2}{*}{\rotatebox{90}{Test~~}}}& DND & 39.37 & 39.37 \\
				\rule[-1ex]{0pt}{3.5ex}
				&SIDD & 35.04 & 39.22 \\				
				\hline
			\end{tabular}
	\end{center}	
	\label{tab:dataset}
\end{table}

\subsection{Ablation Study}
We demonstrate ablation results on the loss terms in \tablename{~\ref{tab:ablation}}. Without the encoder-decoder, using only the restoration term $\mathcal{L}_{res}$ corresponds to the na\"{i}ve blind model, and the performance is inferior to other methods. Without the adversarial loss and the estimation loss, the reconstruction loss is a simple L1 loss. It shows a slightly inferior result compared to using full loss terms because the pixel-wise loss strictly assumes a probability distribution family. As the pixel-wise loss is known to generate blurry results, it tends to see more on structures, where the noise is more likely textures and details. Hence, the noise distribution cannot be well captured by pixel-wise loss terms. By using all loss terms, the proposed method guarantees the performance that surpasses the others. Consistent results on real-noise are presented in \tablename{~\ref{tab:ablation2}}. Note that the experiment is performed with the SIDD validation set in this case, instead of the test set that requires online submissions. In the case of DND, it is tested by submitting the results to the DND site because the validation set is not available.

For further investigation on the framework, we replace the denoiser with FFDNet \cite{FFDNet} architecture and retrain with our framework. The comparisons with the original FFDNet \cite{FFDNet} are shown in \tablename{~\ref{tab:FFDNet}}. There exists noticeable performance gain due to our VDIR framework.
From this result, we can see that the performance gain over FFDNet, shown in Tables \ref{tab:AWGN}, \ref{tab:SIDD}, and \ref{tab:DND}, are not just from the use of more parameters. Specifically, we have the gains over the FFDNet by using the same-sized FFDNet architecture into our VDIR framework, which validates the effectiveness of our variational scheme.
Also, it is noteworthy that the results of FFDNet are non-blind results, whereas our method is tested on blind conditions. Even though they are tested on different conditions, our framework boosts the performance.
In conclusion, our framework does not only boost the performance, but also works as a noise level estimator, making the overall framework to be a blind denoiser, which is very effective.

\begin{table}
	\caption{Ablation study on CBSD68 with $\sigma=30$.}
	\label{tab:ablation}
	\centering
	\begin{tabular}{l|c}
		\hline
		\rule[-1ex]{0pt}{3.5ex}
		Loss term     & PSNR \\
		\hline
		\rule[-1ex]{0pt}{3.5ex}
		$\mathcal{L}_{denoise}$ & 30.47 \\
		\rule[-1ex]{0pt}{3.5ex}
		$\mathcal{L}_{denoise} + \beta D_{KL} + \mathcal{L}_{recon}$ w/o $\mathcal{L}_{adv}$ & 30.56 \\
		\rule[-1ex]{0pt}{3.5ex}
		$\mathcal{L}_{denoise} + \beta D_{KL} + \mathcal{L}_{recon}$ w/ $\mathcal{L}_{adv}$ & 30.64\\
		\hline
	\end{tabular}
\end{table}

\begin{table}
	\caption{Ablation study on real-noise. ``SIDD-val'' means that the comparison is performed with the SIDD validation set.}
	\label{tab:ablation2}
	\centering
	\begin{tabular}{l|c|c}
		\hline
		\rule[-1ex]{0pt}{3.5ex}
		Loss term     &  SIDD val & DND \\
		\hline
		\rule[-1ex]{0pt}{3.5ex}
		$\mathcal{L}_{denoise}$ & 39.11 & 39.47 \\
		\rule[-1ex]{0pt}{3.5ex}
		$\mathcal{L}_{full}$ & 39.29 & 39.63\\
		\hline
	\end{tabular}
\end{table}

\begin{table}[t]
	\caption{The results of the denoiser replaced by FFDNet\cite{FFDNet} on Urban100.}
	\label{tab:FFDNet}
	\centering
	\begin{tabular}{l|c|c}
		\hline
		\rule[-1ex]{0pt}{3.5ex}
		Noise level  &  $\sigma = 30$ & $\sigma = 50$ \\
		\hline
		\rule[-1ex]{0pt}{3.5ex}
		FFDNet \cite{FFDNet}  & 30.53 & 28.05 \\
		\rule[-1ex]{0pt}{3.5ex}
		Ours (FFDNet) & 30.82 & 28.49 \\
		\hline
		\rule[-1ex]{0pt}{3.5ex}
		$\Delta$PSNR & 0.29 & 0.44 \\
		\hline
	\end{tabular}
\end{table}

\section{Conclusion}
We have proposed a novel universal variational approach for image restoration, which can be applied to many image restoration tasks. Concretely, we reformulated the joint distribution of the low-/high-quality images based on the proposed latent variable $\c$, which brings out two inference problems.
Then, we introduced a \emph{variational lower bound} to approximate the inference problem of the latent variable. With our \emph{variational lower bound}, the original inference problem can be simplified and divided into separate sub-problems. Our \emph{variational lower bound} incorporates both the restoration objective and the reconstruction objective, which is a generative model on the low-quality image. Hence, the embedded information of complicated low-quality image manifold plays a role as additional prior information. We have also presented three parameterized CNNs for the inference problem and have shown that our method achieves state-of-the-art performance in several image restoration tasks while requiring moderate model capacity. Furthermore, we also presented more analyses of our framework in the {\em supplemental material} to demonstrate that our method has steerability, which can be applied to practical applications. Our code and some visualized results are available at \url{https://github.com/JWSoh/VDIR}.

%
%
%
%

\ifCLASSOPTIONcaptionsoff
  \newpage
\fi

\end{document}